\documentclass[a4paper,11pt]{article}
\pdfoutput=1
\usepackage{jheppub}
\usepackage{verbatim}
\usepackage{graphicx}
\usepackage{subfigure}
\usepackage{amssymb,amsmath,amsfonts}
\usepackage{tikz}
\usepackage{pgfplots}
\pgfplotsset{compat=default}
\usepackage{tkz-euclide}
\usetkzobj{all}

\def\be{\begin{eqnarray}}
\def\ee{\end{eqnarray}}
\def\rmd{{\rm d}}

\title{
Rigid Holography and \\
Six-Dimensional ${\cal N}=(2,0)$ Theories on AdS$_5\times \mathbb{S}^1$
\\ }

\author[a]{Ofer Aharony,}
\author[a]{\ Micha Berkooz,}
\author[b,c]{\ Soo-Jong Rey}

\affiliation[a]{Department of Particle Physics and Astrophysics \\
Weizmann Institute of Science, Rehovot 7610001 \rm ISRAEL}

\affiliation[b]{School of Physics \& Center for Theoretical Physics\\
          Seoul National University, Seoul 151-747 \rm KOREA}

\affiliation[c]{Center for Theoretical Physics of the Universe \\
Institute for Basic Sciences, Daejeon 305-811 \rm KOREA}

\emailAdd{Ofer.Aharony, Micha.Berkooz@weizmann.ac.il, sjrey@snu.ac.kr}

\abstract{
Field theories on anti-de Sitter (AdS) space 
can be studied by realizing them as low-energy limits of AdS vacua of string/M theory. In an appropriate limit, the field theories decouple from the rest of string/M theory. Since these vacua are dual to conformal field theories, this relates some of the observables of these field theories on anti-de Sitter space to a subsector of the dual conformal field theories. We exemplify this `rigid holography' by studying in detail the six-dimensional ${\cal N}=(2,0)$ $A_{K-1}$ superconformal field theory (SCFT) on $AdS_5\times \mathbb{S}^1$, with equal radii for $AdS_5$ and for $\mathbb{S}^1$. We choose specific boundary conditions preserving sixteen supercharges that arise when this theory is embedded into Type IIB string theory on $AdS_5\times \mathbb{S}^5 / \mathbb{Z}_K$. On $\mathbb{R}^{4,1}\times \mathbb{S}^1$, this six-dimensional theory has a $5(K-1)$-dimensional moduli space, with unbroken five-dimensional $SU(K)$ gauge symmetry at (and only at) the origin. On $AdS_5\times \mathbb{S}^1$, the theory has a $2(K-1)$-dimensional `moduli space' of supersymmetric configurations. We argue that in this case the $SU(K)$ gauge symmetry is unbroken everywhere in the `moduli space' and that this five-dimensional gauge theory is coupled to a four-dimensional theory on the boundary of $AdS_5$ whose coupling constants depend on the `moduli'. This involves non-standard boundary conditions for the gauge fields on $AdS_5$. Near the origin of the `moduli space', the theory on the boundary contains a weakly coupled four-dimensional ${\cal N}=2$ supersymmetric $SU(K)$ gauge theory. We show that this implies large corrections to the metric on the `moduli space'. The embedding in string theory implies that 
the six-dimensional ${\cal N}=(2,0)$ theory on $AdS_5\times \mathbb{S}^1$ with sources on the boundary is a subsector of the large $N$ limit of various four-dimensional ${\cal N}=2$ quiver SCFTs that remains non-trivial in the large $N$ limit. The same subsector appears universally in many different four-dimensional ${\cal N}=2$ SCFTs. We 
also discuss a decoupling limit that leads 
to ${\cal N}=(2,0)$ `little string theories' on $AdS_5\times \mathbb{S}^1$.
}

\preprint{SNUST 15-01, WIS/01/15-JAN-DPPA}

\begin{document}
\maketitle

\section{Introduction and Summary}\label{intro}

In the last few years, the study of supersymmetric theories on curved spaces that preserve (some) supersymmetry has developed many interesting techniques, and has led to many new results. In particular, on a set of compact curved spaces, the exact -- twisted or refined -- partition function and supersymmetric indices can be computed by localization techniques applied to various spacetime dimensions \cite{Pestun:2007rz,Kapustin:2009kz,Kallen:2012va}. Comparatively, supersymmetric field theories on non-compact curved spaces were much less studied (see \cite{Burges:1985qq,Aharony:2010ay,Adams:2011vw,Aharony:2011yc,Aharony:2012jf} for a few examples). 

In this paper,  we present a new approach for studying field theories on anti-de Sitter ($AdS$) space, or on products of $AdS$ space with other compact spaces. This is a specific non-compact, maximally symmetric space of negative curvature on which one can put supersymmetric theories while preserving supersymmetry, and we expect that the study of (supersymmetric) field theories on such spaces will also lead to interesting results about the corresponding quantum field theories. We argue that field theories on $AdS$ can be studied by using the AdS/CFT correspondence \cite{Maldacena:1997re, Gubser:1998bc,Witten:1998qj} as a tool. Conventionally, this correspondence is applied to gravitational theories (string or M theory) on $AdS$ space, which are dual to conformal field theories (CFTs).  However, such gravitational theories can contain non-gravitational field theories at low energies living on branes or singularities filling $AdS$ space (or a product of $AdS$ with another compact space) and, in some cases, there is a decoupling limit of these field theories on $AdS$ space from the rest of string/M theory. We will argue that in such a case the non-gravitational field theory on $AdS$ space is `dual' to a sector of the CFT that is dual to the full string/M theory background. 
Since in the bulk the gravity is decoupled and merely provides a fixed background, we will refer to this approach as `rigid holography'. We shall use the rigid holography to learn about strong interaction dynamics of quantum field theories in an $AdS$ background. This `duality' is special, since only part of all possible observables in the field theory on $AdS$ space are captured by the sector of the dual CFT. In particular, it measures the response to sources on the boundary of $AdS$ space. Nevertheless, it can be used to learn about properties of strongly coupled field theories on $AdS$ space that we would not know how to study by other means. We will see that indeed this can lead to many surprising results.

We will focus on a specific example of `rigid holography' -- studying the six-dimensional ${\cal N}=(2,0)$ $A_{K-1}$ theory on $AdS_5\times \mathbb{S}^1$. This theory is interesting since six is the largest space-time dimension in which we have convincing evidence for the existence of non-trivial quantum field theories. A few aspects of the six-dimensional ${\cal N}=(2,0)$ theory are known, but most aspects remain mysterious. These theories defy a Lagrangian description, so we do not know how to study them directly in traditional approaches. Our goal in this paper is to study these theories by putting them on $AdS_5 \times \mathbb{S}^1$, where the two spaces have equal radii and the full supersymmetry is preserved, and to use the `rigid holography' to uncover interesting lower-dimensional `dual' descriptions for them. Specifically, we will use the AdS/CFT correspondence and Gaiotto's description of class ${\cal S}$ theories to relate the six-dimensional ${\cal N}=(2,0)$ $A_{K-1}$ theory on $AdS_5 \times \mathbb{S}^1$ to a subsector in a certain class of four-dimensional ${\cal N}=2$ SCFTs.


For a quantum field theory on $AdS$ space, unlike the ones on a compact space, specifying the base space is not sufficient to uniquely define the theory. This is because boundary conditions on a timelike hypersurface at asymptotic infinity need to be specified for well-posed Cauchy data. Generally, there can exist different viable choices of boundary conditions, subject to kinematic and dynamic conditions such as unitarity, global symmetry and supersymmetry. There are no systematics for this, and a complete classification of boundary conditions is a daunting problem. In the six-dimensional ${\cal N}=(2,0)$ theory, lacking a Lagrangian formulation, there is no known classification. In this paper, we shall study in detail a specific boundary condition for the six-dimensional ${\cal N}=(2,0)$ $A_{K-1}$ theory on $AdS_5\times \mathbb{S}^1$, which preserves the maximal possible supersymmetry (16 supercharges), in the case when the $AdS_5$ and the $\mathbb{S}^1$ have equal radii.  The reason for studying this specific base space and boundary condition is that it arises when embedding this theory into string theory in various ways, the simplest setup being Type IIB string theory on $AdS_5\times \mathbb{S}^5/\mathbb{Z}_K$ \cite{Kachru:1998ys}, as we will describe in detail below\footnote{More generally, we can extend our analysis to ${\cal N} = (2,0)$ ADE theories, embedded into type IIB string theory on $AdS_5 \times \mathbb{S}^5/\Gamma$, where $\Gamma$ refers to an ADE discrete subgroup of $SU(2) \subset SU(3) \subset SO(6)$. }.
We are forced to use such an embedding because the six-dimensional ${\cal N}=(2,0)$ theories are strongly coupled, and we do not know how to study them directly\footnote{At large $K$, they can be studied by utilizing the AdS/CFT correspondence \cite{Maldacena:1997re, Gubser:1998bc,Witten:1998qj}. In this paper, we will not use this limit, and we will focus on these theories with finite values of $K$.}. By embedding the six-dimensional theories into string theory backgrounds that are dual by the AdS/CFT correspondence to four-dimensional ${\cal N}=2$ SCFTs, we will be able to use the `rigid holography' and the properties of these four-dimensional SCFTs to study the enigmatic six-dimensional theory. In particular, utilizing Gaiotto's description of S-duality, we will identify a subsector of these four-dimensional SCFTs (which decouples in an appropriate limit) that captures properties of the ${\cal N}=(2,0)$ theory on $AdS_5\times \mathbb{S}^1$.

Let us first recall some of the salient properties of the six-dimensional ${\cal N}=(2,0)$ $A_{K-1}$ theory, that will lead us to a naive expectation concerning how these theories should behave on $AdS_5\times \mathbb{S}^1$. On $\mathbb{R}^{5,1}$, this theory has a moduli space of supersymmetric vacua spanned by the expectation value of $5(K-1)$ scalar fields,
\be
{\cal M}[\mathbb{R}^{5,1}] = {(\mathbb{R}^5)^{K-1} \over W_K},
\ee
 where $W_K = \mathcal{S}_K$ is the Weyl group of ${A}_{K-1}$ (acting by permutations on $K$ copies of $\mathbb{R}^5$, before taking a quotient by the `center-of-mass' $\mathbb{R}^5$ \footnote{There are some subtleties with taking this quotient and decoupling the `center-of-mass', but they will not be relevant for the purposes of this paper.}). The low-energy theory at a generic point of this $5(K-1)$-dimensional moduli space contains $(K-1)$ free tensor multiplets of the six-dimensional ${\cal N}=(2,0)$ supersymmetry. At the origin of the moduli space, the theory is at an interacting fixed point. On the moduli space, the theory has BPS-saturated strings of finite tension, and their tension goes to zero precisely at the origin.

 We can also consider the six-dimensional ${\cal N}=(2,0)$ theory on $\mathbb{R}^{4,1}\times \mathbb{S}^1$. The finite radius $R$ of $\mathbb{S}^1$ sets a scale and explicitly breaks the underlying conformal invariance.  At energies below the scale $1/R$, the low-energy dynamics is described by the five-dimensional ${\cal N}=2$ supersymmetric Yang-Mills theory with gauge group $SU(K)$ and $g_{YM}^2 \simeq R$. The compactification does not produce additional scalar fields, so the moduli space of this gauge theory retains the same form, now described by the expectation values of five commuting scalar fields in the adjoint representation of $SU(K)$,
 \be
 {\cal M}[\mathbb{R}^{4,1} \times \mathbb{S}^1] = {(\mathbb{R}^5)^{K-1} \over W_K}.
\label{5dmodulispace}
 \ee
At a generic point in the moduli space, the  $SU(K)$ symmetry is spontaneously broken to $U(1)^{K-1}$. At the origin, the full $SU(K)$ gauge symmetry is restored (W-bosons arising from BPS-saturated strings wrapped around the $\mathbb{S}^1$ become massless), and the low-energy dynamics is described by infrared-free $SU(K)$ vector multiplets. Supersymmetry ensures that the flat metric on (\ref{5dmodulispace}) receives no quantum corrections.

When we put the ${\cal N}=(2,0)$ theory on $AdS_5\times \mathbb{S}^1$ with equal radii $R$ of both spaces and with boundary conditions that follow from the embedding in string theory, it has only a $2(K-1)$-dimensional space of supersymmetric vacua\footnote{From the viewpoint of the construction of this theory using type IIB string theory on an orbifold which is locally $\mathbb{C}^2/\mathbb{Z}_K$, the modes parameterizing the space (\ref{ads5modulispace}) are a subset of the modes which parameterize the moduli space (\ref{5dmodulispace}) for the theory defined on $\mathbb{R}^{4,1}\times \mathbb{S}^1$.}
\be
{\cal M}[AdS_5 \times \mathbb{S}^1] = {(\mathbb{C})^{K-1} \over W_K} .
\label{ads5modulispace}
\ee
 On $AdS$ space, vacua with different expectation values of scalar fields differ by non-normalizable modes of these fields. Each boundary condition describes a different theory, and the relation between them is much weaker than between families of supersymmetric vacua in flat space or in compact space \footnote{For instance, different points on the moduli space \eqref{5dmodulispace} are the same at high energies, while different points on the `moduli space' \eqref{ads5modulispace} are not.}. Nevertheless, we will refer to the space of continuously connected supersymmetric vacua (or, equivalently, supersymmetric boundary conditions in the class we study) as a `moduli space'.

 Note that the $AdS_5$ `moduli space' (\ref{ads5modulispace}) is different from the $\mathbb{R}^{4,1}$ moduli space (\ref{5dmodulispace}).
This poses the following question. How does the six-dimensional ${\cal N}=(2,0)$ theory on $AdS_5 \times \mathbb{S}^1$ behave at different points on the `moduli space' (\ref{ads5modulispace})? Far out on the `moduli space', when the vacuum expectation values of the scalar fields are large in units of $1/R$, we expect the low-energy theory below the scale of these vacuum expectation values to still involve $(K-1)$ tensor multiplets, now living on $AdS_5\times \mathbb{S}^1$. These lead to a five-dimensional $U(1)^{K-1}$ gauge theory. Naively, one would expect that, as for $\mathbb{R}^{4,1} \times \mathbb{S}^1$,  the compactification on $AdS_5 \times \mathbb{S}^1$ will give rise to a five-dimensional supersymmetric Yang-Mills theory with gauge group $SU(K)$ on $AdS_5$, which is unbroken at the origin of the `moduli space' (\ref{ads5modulispace}), and is broken to its Cartan subgroup $U(1)^{K-1}$ at other generic points.

However, what we actually find (at least for the set of boundary conditions we will study) is quite different.
We will argue that the above naive picture cannot be correct since it would imply (when this theory is embedded in
string theory) that the four-dimensional ${\cal N}=2$ SCFT, which is the holographic dual of the string theory background in which we embed our ${\cal N}=(2,0)$ theory, should have an enhanced $SU(K)$ global symmetry for a specific value of its coupling constants, but such an enhancement is not possible. We will argue that the useful description near the origin of the `moduli space' indeed involves a five-dimensional $SU(K)$ gauge theory, but that this gauge theory remains unbroken everywhere in the `moduli space' (\ref{ads5modulispace}), so that it is not directly related to the $U(1)^{K-1}$ gauge theory that we find far out on the `moduli space'. The $SU(K)$ gauge theory on $AdS_5$ that we find is always strongly interacting, and it is coupled to a four-dimensional ${\cal N}=2$ $SU(K)$ gauge theory living at the boundary of $AdS_5$, which becomes weakly coupled near the origin of the `moduli space' (where its coupling constant goes to zero)\footnote{Note that our discussion implies that the $SU(K)$ gauge fields in $AdS_5$ bulk obey a different boundary condition than the usual one. At weak coupling this is impossible \cite{Marolf:2006nd},
but we will argue that at strong coupling it is possible.}. This origin turns out to be infinitely far away in the natural metric. 
We also have additional fields living on the boundary of $AdS_5$; at a specific limit in the parameter space we claim that we have a weakly coupled gauge group $\tilde{G} = SU(2)\times SU(3)\times\cdots \times SU(K)$ living there.

In the rest of this paper, we will explain how this surprising picture arises, by analyzing in detail type IIB string
theory on $AdS_5\times \mathbb{S}^5/\mathbb{Z}_K$, and the limit in which this string theory gives rise (at low energies) to the six-dimensional ${\cal N}=(2,0)$ $A_{K-1}$ theory on $AdS_5\times \mathbb{S}^1$.
We will study in detail the behavior of the four-dimensional ${\cal N}=2$ SCFTs that are dual to this string theory background, using the methods developed by Gaiotto and others. We will show that the dual four-dimensional theories have, near the origin of the `moduli space' of the six-dimensional theory, an S-dual weakly coupled $SU(K)$ gauge theory, and that it is natural to think of this gauge group as living on the boundary of AdS space, and coupling to $SU(K)$ gauge fields in the bulk (as well as to additional fields on the boundary). Alternatively, one can try to describe the theory on $AdS_5$ without having any fields living on the boundary, but then one would have  in the bulk an infinite tower of massless higher spin fields that are dual to composites of the free fields of the four-dimensional $SU(K)$ gauge theory (as well as massive higher spin fields that are dual to chiral primary operators), and this seems quite surprising (moreover, it would simply be dual, by performing an additional AdS/CFT duality on the boundary fields, to our previous description).

We will argue that in their large $N$ limit, the four-dimensional ${\cal N}=2$ theories dual to these string theories have a decoupled sector
(that retains non-trivial correlation functions in the large $N$ limit) that is dual to the ${\cal N}=(2,0)$ theory on $AdS_5\times \mathbb{S}^1$, in the sense of capturing its correlation functions with sources on the boundary. This decoupled sector includes (in one of its descriptions) the gauge theory $\tilde G$ mentioned above. Moreover, we will argue that the same decoupled sector appears in many different four-dimensional ${\cal N}=2$ superconformal theories, that have a bulk string theory description that contains at low energies the ${\cal N}=(2,0)$ theory on $AdS_5\times S^1$, arising either from a $\mathbb{Z}_K$ orbifold in type IIB string theory or from $K$ NS5-branes in type IIA string theory.

Our description leads to a boundary condition coupling the $A_{K-1}$ ${\cal N}=(2,0)$ theory to a boundary theory that contains $O(K^3)$ degrees of freedom (in particular this is the number of gauge fields in ${\tilde G}$), and it is interesting to speculate whether there is any relation to the counting of the bulk degrees of freedom of the $A_{K-1}$ theory (that also scales as $O(K^3))$. In order to study this, we look at the finite temperature thermodynamics. 
We find that the two factors of $K^3$ appear in different ways at finite temperature, so there does not seem to be any direct relation between them.

A different low-energy limit of the string theory backgrounds we discuss gives rise to the ${\cal N}=(2,0)$ `little string theory' on $AdS_5\times \mathbb{S}^1$. This theory is even more mysterious than the ${\cal N}=(2,0)$ SCFT, but we will find that it also arises from a decoupled sector of four-dimensional ${\cal N}=2$ SCFTs, in a slightly different decoupling limit. Studying the general class of
four-dimensional ${\cal N}=2$ SCFTs that have such a decoupled sector suggests that there may exist novel `little string theories' with ${\cal N}=(2,0)$ supersymmetry that are not yet known, at least on $AdS_5\times \mathbb{S}^1$.

Because we focus in this paper on a specific example of `rigid holography', we begin by introducing this example in detail. In section \ref{basicsec} we review the four-dimensional ${\cal N}=2$ superconformal quiver gauge theories of $\widehat{A}_{K-1}$ type and their holographic dual type IIB string theory on $AdS_5 \times \mathbb{S}^5/\mathbb{Z}_K$. In section \ref{section3} we discuss the singular limit of these four-dimensional SCFTs in which $(K-1)$ gauge coupling constants are taken to be infinitely large; this singular limit is related to the six-dimensional ${\cal N}=(2,0)$ $A_{K-1}$ theories on $AdS_5\times \mathbb{S}^1$. In section \ref{section4} we discuss the same singular limit in the bulk of $AdS_5$, and we argue that it leads to the picture of the six-dimensional theories on this space that we discussed above. In section \ref{reverse}, we move back and give a general review of `rigid holography', and of decoupling limits giving field theories on $AdS$ space. As this issue was not systematically explored in the literature, and as it may have various applications, we discuss it in a broader context, and describe several additional examples. In section \ref{universality} we return to the ${\cal N}=(2,0)$ $A_{K-1}$ theories on $AdS_5\times \mathbb{S}^1$, and we show that they can be embedded into string theories in many different ways, but that `rigid holography' gives a universal `dual' description for them. In section \ref{section7} we discuss the thermodynamics of these theories. In section \ref{lst} we use similar methods to study `little string theories' on $AdS_5\times \mathbb{S}^1$. Finally, in section \ref{section9}, we discuss further issues and outstanding open questions, some of which we are currently investigating. Various technical materials are relegated to the appendices to make the paper self-contained.

\section{Four-dimensional ${\cal N}=2$ Quiver SCFTs and their Gravity Duals}\label{basicsec}

In this section we recapitulate aspects of the AdS/CFT correspondence between the Type IIB string theory on $AdS_5\times \mathbb{S}^5/\mathbb{Z}_K$ and four-dimensional ${\cal N}=2$ superconformal quiver gauge theory, and bring out puzzles associated with the twisted sector. In subsection \ref{basicduality}, we recall the basic setup. In subsections \ref{orbifoldpoint} and \ref{singularpoint}, we focus on the perturbative orbifold limit and on the singular orbifold limit, respectively. In subsection \ref{singularlimit}, we present the puzzle that the naive application of what (little) is known about the six-dimensional ${\cal N}=(2,0)$ theory compactified on a circle seems to be in conflict with exact statements on the four-dimensional ${\cal N}=2$ superconformal quiver gauge theories.   

\begin{figure}[tbp]
\centering
	\includegraphics[angle=0,width=10cm]{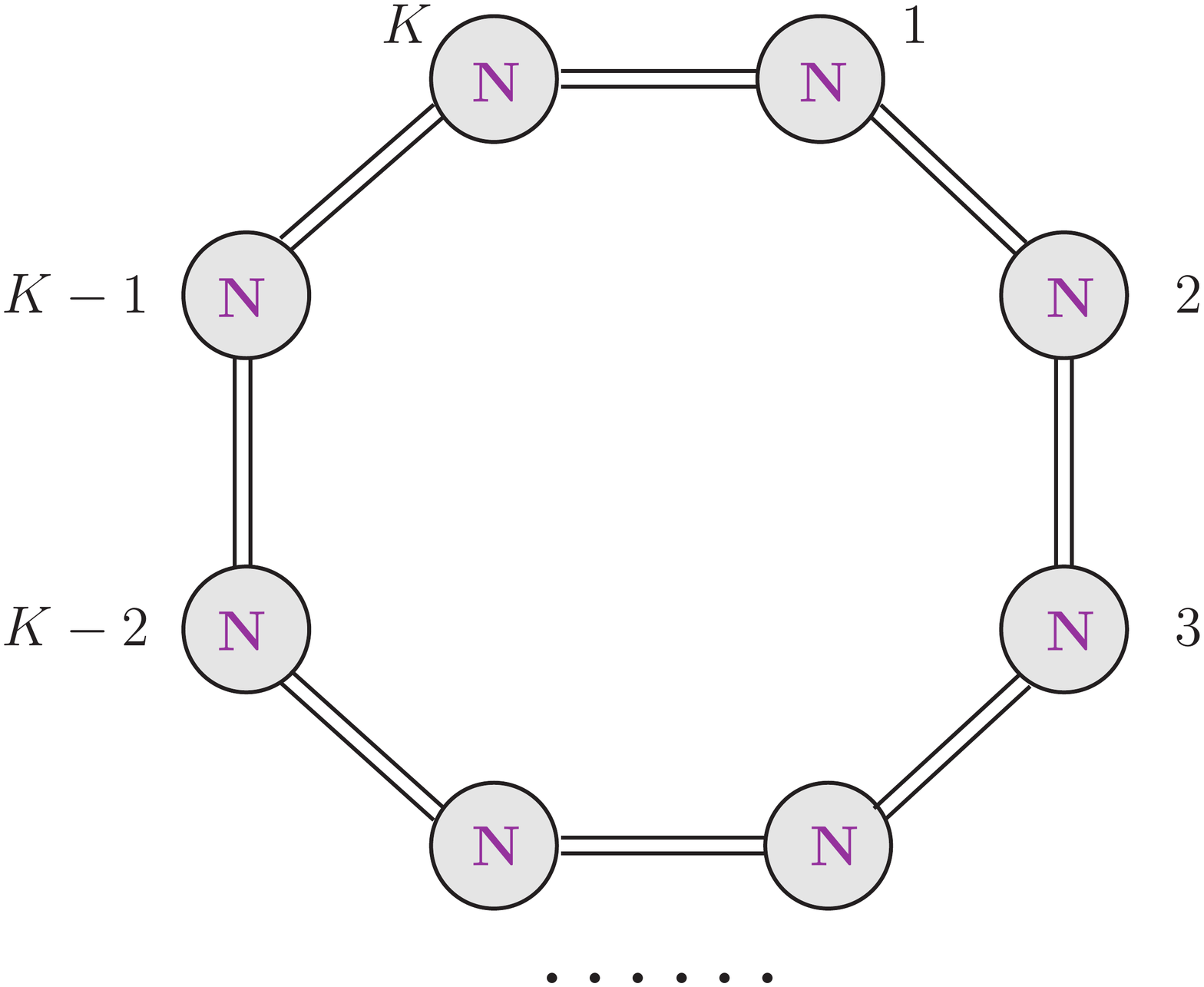}
\caption{\sl The quiver diagram of $\widehat{A}_{K-1}$. In the dual four-dimensional ${\cal N}=2$ SCFT, each node (grey blob) describes a ${\cal N}=2$ vector multiplet in the adjoint representation of $SU(N)$, while each link (double line) describes a ${\cal N}=2$ hypermultiplet in the bifundamental representation of two adjacent $SU(N) \otimes SU(N)$ groups. The global symmetry includes the R-symmetry and a $[U(1)]^K$ coming from the hypermultiplets on each link.}
\label{fig:circular-quiver}
\end{figure}

\subsection{The Basic Duality}
\label{basicduality}
The AdS/CFT correspondence we shall focus on is the duality \cite{Kachru:1998ys} between the Type IIB string theory on  $AdS_5\times \mathbb{S}^5/\mathbb{Z}_K$ and the four-dimensional ${\cal N}=2$ SCFTs with $\widehat{A}_{K-1}$ circular quiver, consisting of vector multiplets of gauge group $G=SU(N)^K$
%
%
and bifundamental hypermultiplets for each adjacent pair of gauge groups (see figure \ref{fig:circular-quiver}).
On the $AdS_5\times \mathbb{S}^5/\mathbb{Z}_k$ side, if one views the $\mathbb{S}^5$ as a fixed radius hypersurface in $\mathbb{C}^3$, then the $\mathbb{Z}_K$ orbifolding acts as $(z_1,z_2,z_3)\rightarrow ( \exp(2\pi i / K) \cdot z_1, \exp(-2\pi i / K)\cdot z_2,z_3)$ \cite{Douglas:1996sw}.
This action leaves invariant the circle $z_1=z_2=0$,
and locally, near this $\mathbb{S}^1$, the orbifold is similar to $\mathbb{C}^2/\mathbb{Z}_K$, the orbifold in flat space that preserves half of the supersymmetries. This duality can be understood, in particular, by considering the low-energy limit of D3-branes on a $\mathbb{C}^2/\mathbb{Z}_K$ orbifold. The various gauge group factors in $G$ then correspond to
`fractional branes' that can move when the branes are on top of the orbifold singularity. 

In four-dimensional ${\cal N}=2$ SCFTs, the bosonic symmetries include an $SO(4,2)$ conformal symmetry and $SU(2)_R \times U(1)_R$ R-symmetries. Additionally, in our case there is the $U(2)$ or $U(1)^K$ global symmetry carried by the hypermultiplets for $K = 2$ or $K > 2$, respectively\footnote{The extra global symmetry in the $K=2$ case arises because we have two hypermultiplets in the same representation.}.
In type IIB string theory, some of these bosonic symmetries are realized geometrically. The orbifolding retains the $AdS_5$ isometry $SO(4,2)$ but breaks the $SO(6)_R$ isometry of the covering $\mathbb{S}^5$ to $SU(2)_R\times U(1)_R \times SU(2)$ for $K=2$, and to $SU(2)_R\times U(1)_R\times U(1)$ for $K > 2$. 
Note that the $U(1)_R$ symmetry arises from an isometry of the $\mathbb{S}^1 \subset \mathbb{S}^5$ that is by construction fixed under the orbifold. The supercharges are charged under this $U(1)_R$ and hence rotated under a shift along the $\mathbb{S}^1$.

\begin{figure}[tbp]
\vspace*{-2cm}	
\includegraphics[angle=0,width=15cm]{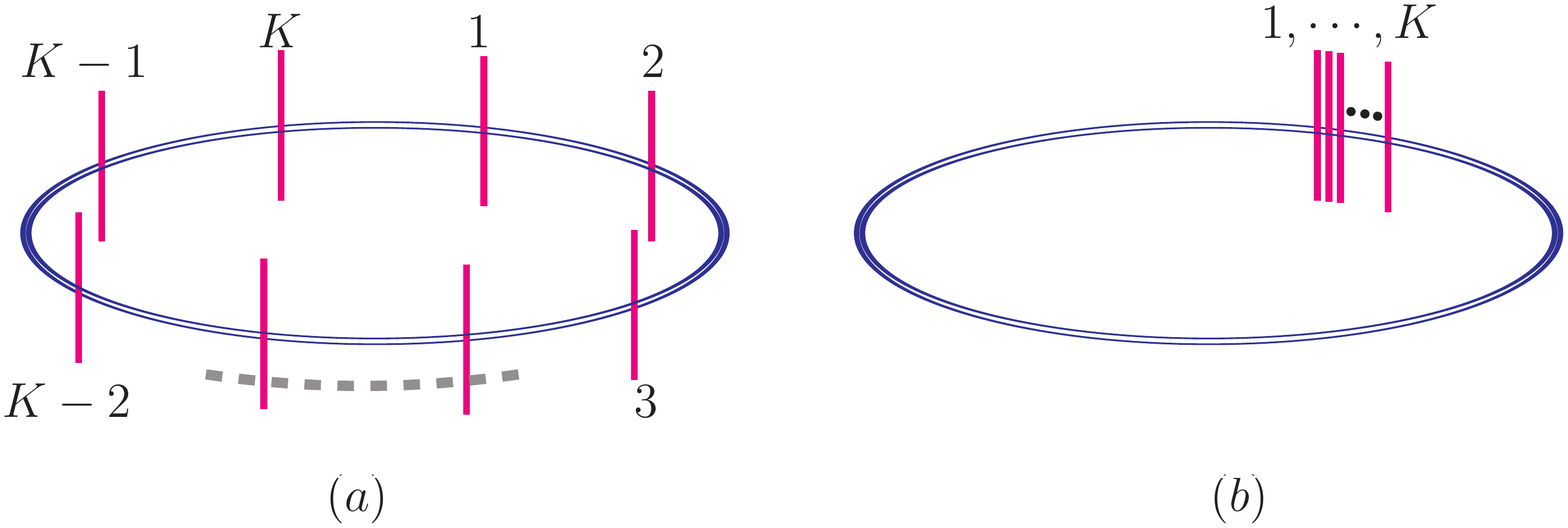}
\vspace*{-2cm}
\caption{\sl The Type IIA brane configuration of the four-dimensional quiver theory, at the perturbative orbifold point (a) and at the singular orbifold point (b) that we will discuss below. The red lines are NS5-branes, and the blue lines are D4-branes.}
\label{fig:brane-config}
\end{figure}

Each node of the circular quiver is an ${\cal N}=2$ supersymmetric $SU(N)$ gauge theory with $2N$ hypermultiplets, and the beta function vanishes exactly. So the complex coupling constants for each gauge group in $G$:
\be
\tau_a \equiv \frac{\theta_a}{2\pi} + i \frac{4\pi }{g_{{\rm YM},a}^2} \qquad (a = 1, \cdots, K)
\label{complexcoupling}
\ee
%
are exactly marginal deformations of the quiver gauge theory, such that there is a SCFT for every value of these couplings. In fact, these are the only marginal deformations that preserve the ${\cal N}=2$ super(conformal) symmetry.  The global structure of this space of coupling constants was studied using a Type IIA brane construction of these field theories \cite{Witten:1997sc} (see figure \ref{fig:brane-config}). There, it was shown that, taking various dualities into account, the space of these couplings is the complex structure of a $K$-punctured torus $\mathbb{T}^2_K$. 
Denote the complex structure of the torus $\mathbb{T}^2$ as $\tau_0$ and the local coordinate of the $a$-th puncture as $\omega_a$.
The complex coupling constant of the $a$-th quiver gauge field is then given by
\begin{equation}
\tau_a \quad \leftrightarrow \quad 
\omega_a - \omega_{a-1} \qquad (a=1, \cdots, K),
\label{fortaua}
\end{equation}
where $\omega_{0}$ is to be understood as $\omega_K$.
Being periodic on the torus, 
\be
\omega_a \simeq \omega_a + \mathbb{Z} \simeq \omega_a + \tau_0\mathbb{Z},
\label{omega}
\ee
and we can use these identifications in (\ref{fortaua}) to make $0 \leq {\rm Im}(\omega_a) \leq {\rm Im}(\tau_0)$, and  also $0 \leq {\rm Im}(\tau_a) \leq {\rm Im}(\tau_0)$.
The quiver being circular, the complex gauge coupling constants \eqref{fortaua} obey:
\begin{equation}
\hskip-1.9cm \sum_{a=1}^{K} \tau_a = \tau_1 + \cdots + \tau_K = \tau_0.
\label{sum}
\end{equation}
For later considerations, we shall take the $K$ independent coupling parameters to be $\tau_0, \tau_1, \cdots, \tau_{K-1}$.

On the Type IIB string theory side, by holography, we can also identify the space of exactly marginal deformations. Exactly marginal deformations in the SCFT map to massless scalar fields in the string theory that have no potential.
The coupling $\tau_0$ is mapped, as in the four-dimensional ${\cal N}=4$ super Yang-Mills theory, to the complex scalar field of the Type IIB string, consisting of the dilaton $\Phi$ and the Ramond-Ramond (R-R) 0-form $C_0$:
\be
\tau_0 \quad \leftrightarrow \quad {C_0 \over 2 \pi} + i {4 \pi \over e^{\Phi}}.
\ee
The other $(K-1)$ couplings are mapped to massless fields that arise from the twisted sector of the $\mathbb{Z}_K$ orbifold. 

In ten-dimensional Minkowski space, the massless modes of Type IIB string theory living at the fixed locus of the $\mathbb{C}^2/\mathbb{Z}_K$ orbifold are $(K-1)$ tensor multiplets of the six-dimensional ${\cal N}=(2,0)$ supersymmetry. Each such tensor multiplet contains a self-dual 2-form field, four symplectic Majorana fermions and five real scalars, giving rise (locally) to an $\mathbb{R}^{5(K-1)}/S_K$ moduli space. These moduli are associated with the 
$(K-1)$ non-trivial homology two-cycles 
that can be blown up from this orbifold singularity. $3(K-1)$ of the scalar fields correspond to the blow-up modes of these 2-cycles,
and $2(K-1)$ correspond to integrals of the NS-NS and R-R 2-forms $B_2, C_2$  over these homology 2-cycles
\cite{Aspinwall:1994ev}. When this orbifold is embedded in $AdS_5\times \mathbb{S}^5/\mathbb{Z}_K$, the local physics is the same but the global structure at the curvature scale is different. The fields above generally couple non-minimally to the background curvature, and acquire masses of order the inverse radius. The $3(K-1)$ blow-up modes are triplets of the $SU(2)_R$ symmetry, and their zero modes on the $\mathbb{S}^1$ become tachyonic fields on $AdS_5$, at the
Breitenlohner-Freedman bound \cite{Breitenlohner:1982jf}\footnote{We review the Kaluza-Klein spectrum of the twisted sector of type IIB string theory on $AdS_5\times \mathbb{S}^5/\mathbb{Z}_K$ in Appendix \ref{twistedspectrum}.}.
On the other hand, the zero modes on the $\mathbb{S}^1$ of the $2(K-1)$ scalars coming from the periods of the NS-NS and R-R two-forms over the homology two-cycles remain massless, and are holographically dual to the $(K-1)$ additional complex exactly marginal deformations of the four-dimensional ${\cal N}=2$ SCFT (\ref{complexcoupling}): one can choose a basis of cycles $\Sigma_a$ such that
\be
\tau_a \quad \leftrightarrow \quad \oint_{\Sigma_a} {C_2 \over 2 \pi}   + \tau_0 \oint_{\Sigma_a} {B_2 \over 2 \pi} \qquad \qquad (a = 1, \cdots, K-1).
\ee
With the normalization specified, the integrated 2-form potentials are periodic with period one.
%
%

The structure of global symmetries will play a particularly important role in what follows. The full global symmetry group of the SCFTs (for $K>2$) contains 
the $U(1)^K$ global symmetry acting on each of the $K$ bifundamental hypermultiplets. In the string theory description, this global symmetry ought to be mapped to a $U(1)^K$ gauge symmetry on $AdS_5$. As we discussed at the beginning of this section, one of these $U(1)$ factors arises geometrically as the Abelian isometry of $\mathbb{S}^5/\mathbb{Z}_K$; this `center-of-mass' $U(1)$ rotates all of the hypermultiplets simultaneously. The other `relative' $(K-1)$ $U(1)$'s arise from the twisted sectors of the orbifold. As explained above, there are $(K-1)$ massless tensor multiplets living at the orbifold fixed point, $AdS_5\times \mathbb{S}^1$. Each of these multiplets contains a self-dual 2-form field. The Kaluza-Klein reduction of this field on the $\mathbb{S}^1$ contains a massless vector field on $AdS_5$, which we identify with one of the $(K-1)$ relative $U(1)$ global symmetries in the ${\cal N}=2$ SCFT with $\widehat{A}_{K-1}$ quiver gauge group.

\subsection{Orbifold Point}
\label{orbifoldpoint}

Let us focus next on the orbifold point in parameter space. This is the easiest SCFT to analyze. 
In this case, both the string theory background and the field theory parameters manifest a $\mathbb{Z}_K$ symmetry, and 
\be\label{orbifoldvalue}
\tau_1 = \tau_2 = \cdots = \tau_{K} = {\tau_0 \over K} = \frac{1}{K} \left({\theta_0 \over 2 \pi} + i {4 \pi \over g^2_0} \right).
\ee
This is the point in parameter space which corresponds to applying standard worldsheet orbifold techniques to D3-branes. At this point, the mapping of the gauge theory parameters $N$ and $\tau_0$ to string theory background fields is identical to the mapping in the underlying duality between the four-dimensional ${\cal N}=4$ super Yang-Mills theory and the Type IIB string theory on $AdS_5\times \mathbb{S}^5$. In particular, $N$ is the flux of the R-R 5-form on $\mathbb{S}^5/\mathbb{Z}_K$. When ${\rm Im}(\tau_0) \gg N$, the field theory is weakly coupled and perturbation theory is reliable, but the string theory lives on a highly curved bulk space. When $1 \ll {\rm Im}(\tau_0) \ll N$, the field theory has large $N$ and large 't Hooft coupling $\lambda = (4 \pi N/ \rm{Im} (\tau_0))$, but the bulk string theory is weakly coupled and weakly curved, providing a useful perturbative description.

At the orbifold point, we can compute some objects reliably. For example, on the field theory side, at least at weak coupling, all the operators charged under the 'relative' $U(1)^{K-1}$ global
symmetries have large scaling dimensions. For instance, such operators arise by taking the determinant of one of the
scalars in a bi-fundamental hypermultiplet, giving an operator of classical dimension $N$ \footnote{Each hypermultiplet contains two complex scalars, $Q$ and $\widetilde Q$, such that $Q$ is in the same gauge group representation as ${\widetilde Q}^*$. One can form both chiral operators like $\det(Q)$, and non-chiral operators by replacing some of the $Q$'s in the determinant by ${\widetilde Q}^*$'s.}. On the string theory
side, charged states arise from D3-branes wrapped on one of the homology 2-cycles of the orbifold and on the
$\mathbb{S}^1$. Even though the homology 2-cycle has a zero area, these D3-branes acquire a mass from the NS-NS two-form potential on the vanishing 2-cycle. At the orbifold point, where the worldsheet conformal field theory description is well-behaved, the values of the 2-form fields are \cite{Aspinwall:1994ev}:
\be
\oint_{\Sigma_a} {C_2 \over 2 \pi} = 0,\quad \oint_{\Sigma_a} {B_2 \over 2 \pi} = {1 \over K} \quad \mbox{for \ every} \quad a= 1, \cdots, K-1,
\ee
realizing (\ref{orbifoldvalue}).
So, the wrapped D3-brane has $1/K$ of the tension of a D-string. The mass of these non-BPS objects at the orbifold point, in units of the $AdS_5$ radius, is of order $N/K\sqrt{\lambda}$ for large $\lambda$. The fact that this is much smaller than $N$ at strong coupling implies that some of the corresponding charged operators acquire large negative anomalous dimensions at strong coupling. This is quite surprising, but it is a straightforward logical consequence of our discussion \footnote{We do not know of any other examples where such large negative anomalous dimensions arise at strong coupling, even for operators with dimensions of order $N$, for which the large $N$ corrections are under less control. However, this is what we find in our example. It would be interesting to find further examples of this phenomenon.}.


\subsection{Singular Point}
\label{singularpoint}

Our primary interest is to study what happens in the theory when we move away from the orbifold point and, in particular,
when we go to the singular point where all the NS-NS and R-R 2-forms integrated over the vanishing homology 2-cycles associated with the orbifold vanish:
\be
\oint_{\Sigma_a} {B_2 \over 2 \pi}, \qquad \oint_{\Sigma_a} {C_2 \over 2 \pi} \quad \rightarrow \quad 0 \quad \mbox{for \ all} \quad a=1, \cdots, K-1.
\ee
%
%
%

Let us first review this singular limit for the flat space orbifold $\mathbb{C}^2/\mathbb{Z}_K$ \cite{Aspinwall:1994ev}.
When the blow-up modes of a 2-cycle $\Sigma_a\ (a=1, \cdots, K-1)$ vanish, the D3-brane wrapped on
that 2-cycle of the orbifold is a BPS string state. It has a tension set by the values of the NS-NS and R-R 2-forms on the 2-cycle, $T_{D3} \propto |\oint_{\Sigma_a} (C_2 + \tau_0 B_2)| / \ell_{\rm st}^2$. In particular, when both 2-forms go to zero, it becomes tensionless, so the resulting low-energy theory at the singular point of the orbifold is a conformally invariant theory that involves these {\sl tensionless strings} as an integral part of the spectrum. Despite such nonlocal excitations, string dualities and other arguments indicate that the low-energy theory at this singular point is a {\sl local} and conformally invariant quantum field theory, known as the six-dimensional $A_{K-1}$ ${\cal N}=(2,0)$ theory. The same theory arises at low energies also on $K$ coincident NS5-branes in Type IIA string theory, or on $K$ coincident M5-branes in M theory. Note that the resolution of $\mathbb{C}^2/\mathbb{Z}_K$ to $(K-1)$ $\mathbb{S}^2$'s has intersections given by the $A_{K-1}$ Dynkin diagram. The six-dimensional ${\cal N}=(2,0)$ theory has a moduli space $({\mathbb{R}^5)^{K-1}}/S_K$, uniquely determined by the supersymmetry, and the effective low-energy
theory on the moduli space contains $(K-1)$ massless tensor multiplets, as we found at the orbifold point. In the M theory picture,  the BPS strings, whose tension goes to zero at the origin of the moduli space and which can be identified with the wrapped D3-branes described above, are realized as M2-branes stretched between a pair of M5-branes. Note that, in the string theory construction, the six-dimensional ${\cal N}=(2,0)$ theory provides a good description of the theory at energies much lower than the string scale. In particular, the wrapped D3-branes can only be thought of as part of the six-dimensional local quantum field theory when the two-form fields are small enough that the wrapped D3-brane tension is much smaller than the fundamental string tension. For large values of $K$, the ${\cal N}=(2,0)$ $A_{K-1}$ theory
can be studied using the AdS/CFT correspondence \cite{Maldacena:1997re}. For finite values of $K$, the values of its anomalies are known \cite{Harvey:1998bx} and some information
protected by supersymmetry can be derived by various methods, but not much else is known.

What happens when we take the same singular limit in type IIB string theory on $AdS_5\times \mathbb{S}^5/\mathbb{Z}_K$ ? (The situation for $K=2$ was previously studied in \cite{Rey:2010ry} and also in 
\cite{Gadde:2010zi}.)
First, we can identify what this limit corresponds to in the space of exactly marginal deformations that we
described above. As already mentioned, the $2(K-1)$ values of the 2-form potentials correspond to the relative positions
of the $K$ marked points on the torus $\mathbb{T}^2_K$ in the description of \cite{Witten:1997sc}. It is clear that taking the period integral of these 2-forms to zero corresponds to bringing all $K$ marked points on the torus together\footnote{We can also bring $k < K$ points together, corresponding to taking only some of the two-forms to zero, and obtaining at low energies an $A_{k-1}$ theory in the bulk.}. This is the singular limit we are interested in. From the viewpoint of the four-dimensional ${\cal N}=2$ SCFT, one can think of this limit as having $(K-1)$ of the coupling constants go to infinity, such that $(K-1)$ of the $\tau_a$ go to zero while one $\tau_a$ remains finite (and is nearly equal to $\tau_0$):
\be
\tau_1, \cdots, \tau_{K-1} \quad \rightarrow \quad 0 \quad \mbox{while} \quad \tau_0 = \mbox{finite}.
\ee
%
Actually, there is a variety of ways of taking this infinite coupling limit. For example, we can take the infinite coupling limit such that the ratios among the coupling constants are finite, or we can take some to infinity faster. We will use later some of these different possibilities \footnote{Note that this limit 
is very different from the opposite limit, where the $K$ coupling constants becomes weak. In that limit, while the ratios among the coupling constants may be similar to the singular limit we consider, the corresponding $\tau_a$'s, and also $\tau_0$, all go to infinity, and the four-dimensional ${\cal N}=2$ SCFT is weakly coupled and can be treated within perturbation theory. In this weak coupling limit, the entire dual bulk description becomes highly curved.}.

\subsection{The Puzzle at the Singular Limit}
\label{singularlimit}

When we take the singular limit in the bulk, we expect that the local description of the singularity is the same as in flat space. Of course, the global structure modifies physics at the scale $1/R$, where $R$ is the equal value of the radii of the $AdS_5$, the covering $\mathbb{S}^5$ and the $\mathbb{S}^1$. As we are assuming that the 't Hooft coupling is large, this mass scale is much smaller than the string mass scale and, in particular, can be kept fixed while the string mass scale is taken to infinity. So, we expect to be able to describe the singular limit of type IIB string theory on $AdS_5\times \mathbb{S}^5/\mathbb{Z}_K$ in terms of the six-dimensional ${\cal N}=(2,0)$ $A_{K-1}$ theory on $AdS_5\times \mathbb{S}^1$, weakly coupled (with couplings of order $1/N$) to the rest of the string theory. This description should be valid whenever the tension of the wrapped D3-branes is much smaller than the fundamental string tension, and the corresponding region of the `moduli space' of type IIB string theory on $AdS_5\times \mathbb{S}^5/\mathbb{Z}_K$ looks like \eqref{ads5modulispace}.

Very little is known about the ${\cal N}=(2,0)$ $A_{K-1}$ theory for finite values of $K$, but one of the few things that is known
is that, when this theory is compactified on $\mathbb{R}^{4,1} \times \mathbb{S}^1$, the resulting low-energy theory (below the scale of the
inverse radius of the circle) is a five-dimensional maximally supersymmetric Yang-Mills theory with gauge group $SU(K)$, with a Yang-Mills coupling constant squared proportional to the radius of the $\mathbb{S}^1$
\cite{Rozali:1997cb}.
However, we will see that applying this to the case at hand leads to a puzzle. 

Naively, one might think that,  since our six-dimensional ${\cal N}=(2,0)$ $A_{K-1}$ theory is compactified on $\mathbb{S}^1$, the low-energy dynamics is described by a massless five-dimensional $SU(K)$ gauge theory living on the $AdS_5$ space. The six-dimensional conformal invariance  implies that the five-dimensional gauge theory sits at {\sl the origin of the `moduli space'} \eqref{ads5modulispace}, where all integrals of two-form fields over the $(K-1)$ 2-cycles vanish. Moreover, one expects that this $SU(K)$ gauge symmetry is the non-Abelian completion of the $U(1)^{K-1}$ gauge symmetry in $AdS_5$, which is visible far out on the `moduli space'. Such an $SU(K)$ gauge theory would be strongly coupled since its coupling constant is at the same scale as the $AdS_5$ radius, which sets the minimal energy scale, but one may still suppose that there are states in the bulk coming from massless $SU(K)$ gauge fields. Since gauge symmetry in the bulk corresponds to global symmetry in the dual SCFT, this expectation is translated to the assertion that the four-dimensional dual ${\cal N}=2$ quiver SCFT would have an $SU(K)$ global symmetry at the singular point in its parameter space, enhancing the manifest $U(1)^{K-1}$ global symmetry. This, however, is not possible. 

This is not possible because four-dimensional ${\cal N}=2$ SCFTs cannot have enhanced global symmetries as exactly marginal couplings are continuously varied, without having also enhanced higher-spin symmetries (see also \cite{Beem:2013sza,Beem:2014zpa}). We can see this by examining the list of allowed multiplets of the four-dimensional ${\cal N}=2$ superconformal algebra \cite{Dobrev:1985vh,Dobrev:1985qz,Dobrev:1985qv,Minwalla:1997ka,Dolan:2002zh}, in which there is no superconformal multiplet including a vector operator with a dimension that continuously goes down to the conserved current dimension $\Delta=3$, without containing also additional conserved currents of higher spins\footnote{In our case,  we know that the $U(1)^{K-1}$ gauge fields live in a standard vector multiplet that does not contain conserved higher spin fields, so clearly any extra fields that enhance the symmetry to $SU(K)$ should not come with such higher spin fields either. On the other hand, extra global symmetries together with extra
conserved higher-spin currents do arise in non-interacting limits of four-dimensional ${\cal N}=2$ SCFTs.}.

Indeed, there are several loose points in the expectation that such an enhanced $SU(K)$ gauge symmetry should arise. First, even if such an enhanced symmetry were to appear when the five-dimensional $SU(K)$ gauge theory is weakly coupled, $R_{\mathbb{S}^1} \ll R_{AdS_5}$, supersymmetric field theories on $AdS_5$ can undergo phase transitions as the coupling constants are dialed (see \cite{Aharony:2012jf} for a recent discussion), and there is no argument that the enhanced symmetry would survive at the strong coupling regime, where $R_{\mathbb{S}^1} \simeq R_{AdS_5}$. Note that the D3-branes wrapped on the orbifold blow-up cycles $\Sigma_a$ and on $\mathbb{S}^1$, corresponding to the putative W-bosons of $SU(K)$, carry $[U(1)]^{K-1}$ charges, but they are not BPS states on $AdS_5$, so they do not have to become massless even at the origin of the `moduli space'~\footnote{On the other hand, the D3-branes wrapped just on $\Sigma_a$ do give BPS strings also on $AdS_5$, with the same tension as in flat space, and they do become tensionless at the origin.}.  Second, the $\mathbb{S}^1$ compactification of the six-dimensional ${\cal N}=(2,0)$ theory onto $AdS_5$ space is different from the one in flat space, as we need to provide more data in the form of boundary conditions for the fields as well as for the supercharges. This is related to the fact that the isometry of the $\mathbb{S}^1$ is part of the R-symmetry group of the theory in $AdS_5$.
In the next sections we will describe in detail what actually happens around the origin of the `moduli space'.


\section{The Singular Limit: Four-dimensional ${\cal N}=2$ SCFT Perspective}
\label{section3}

We have argued that the naive expectation, that the $U(1)^{K-1}$ gauge symmetry of the $AdS_5\times \mathbb{S}^5/\mathbb{Z}_K$ string theory ought to be enhanced to $SU(K)$ at the origin of the `moduli space', leading to an enhanced $SU(K)$ global symmetry in the four-dimensional SCFT, cannot possibly be correct. So, what do we get at this singular point? To reconcile the bulk intuition with the exact field theory obstruction, we can study what happens in this limit in the four-dimensional ${\cal N}=2$ SCFT by using methods that were recently developed by Gaiotto and others for the study of such theories that can be described as compactified M5-branes \cite{Gaiotto:2009we,Gaiotto:2009hg,Chacaltana:2010ks,Chacaltana:2012zy}. A large class of  four-dimensional ${\cal N}=2$ SCFTs, called class ${\cal S}$, can be described using six-dimensional ${\cal N}=(2,0)$ SCFTs compactified on Riemann surfaces with punctures (called `UV curves'), and such a description is useful for understanding strong-weak coupling S-dualities and other symmetry properties (see \cite{Tachikawa:2013kta,Gaiotto:2014bja} for reviews). 

\begin{figure}[tbp]
\centering
	\includegraphics[angle=0,width=10cm]{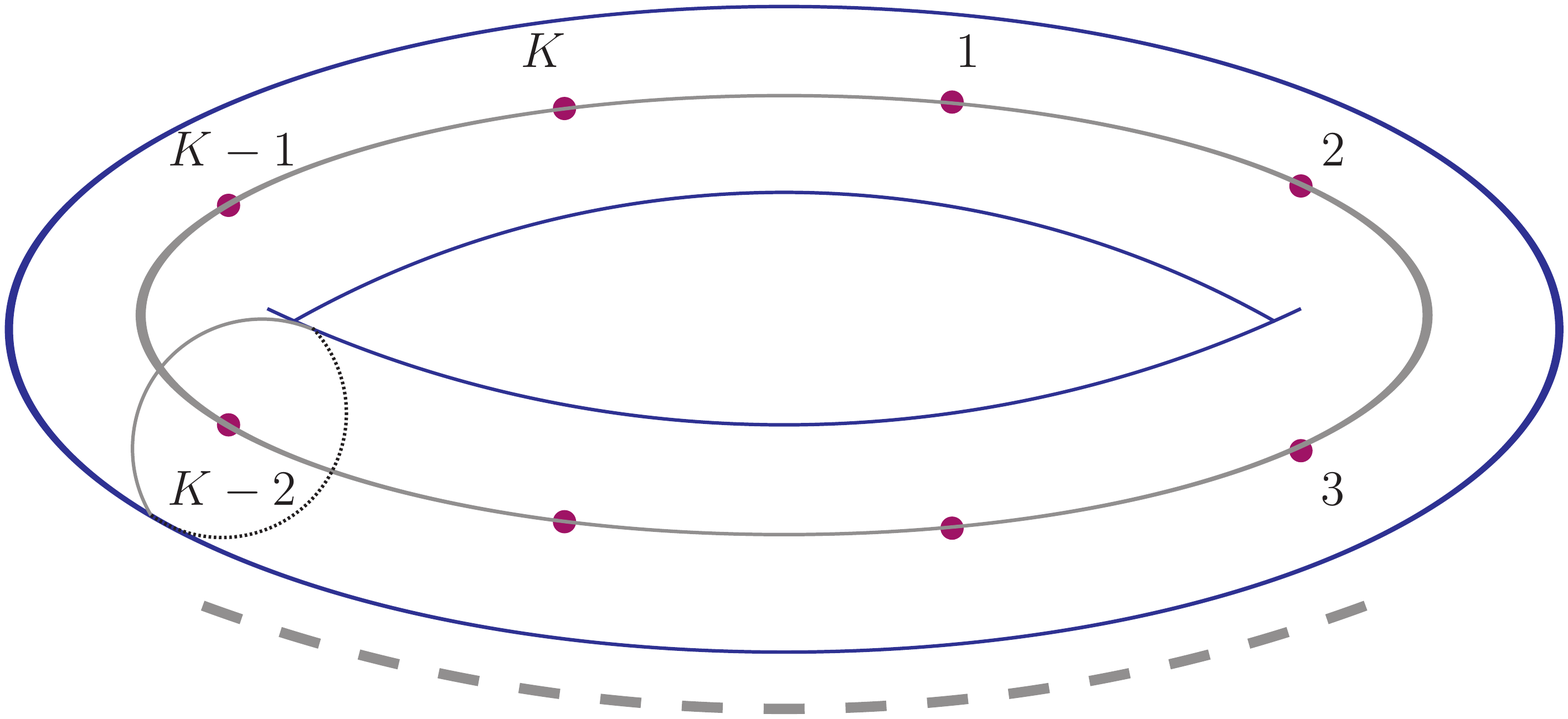}
\caption{\sl The UV curve of the four-dimensional ${\cal N}=2$ SCFT with $\widehat{A}_{K-1}$ quiver gauge group.}
\label{fig:circular-quiver-orbifold}
\end{figure}

The four-dimensional ${\cal N}=2$ SCFTs we discuss here can also be described as such a compactification \cite{Witten:1997sc}: they arise from the six-dimensional ${\cal N}=(2,0)$ $A_{N-1}$
SCFTs compactified on a torus, with modular parameter $\tau_0$, and with $K$ punctures on this torus, see figure \ref{fig:circular-quiver-orbifold}.
It is now known that many different types of punctures are possible in compactified six-dimensional theories,
which differ by the amount of global symmetry that they give rise to. From the consideration of the global symmetries of the four-dimensional ${\cal N}=2$ SCFTs described above, it is clear that in our case each puncture is associated with a $U(1)$ global symmetry, so that the punctures are {\sl minimal regular punctures} in the standard terminology \cite{Chacaltana:2010ks,Tachikawa:2013kta}. Beware that this six-dimensional ${\cal N}=(2, 0)$ $A_{N-1}$ theory should not be confused with the six-dimensional ${\cal N}=(2,0)$ $A_{K-1}$ theory that we have in the bulk dual description\footnote{Rather, as we will see below, some sector of the former theory, dimensionally reduced on a punctured torus times $\mathbb{R}^{3,1}$, is holographically `dual' to the latter theory on $AdS_5 \times \mathbb{S}^1$.}.

\begin{figure}[tbp]
\centering
	\includegraphics[angle=-90,width=10cm]{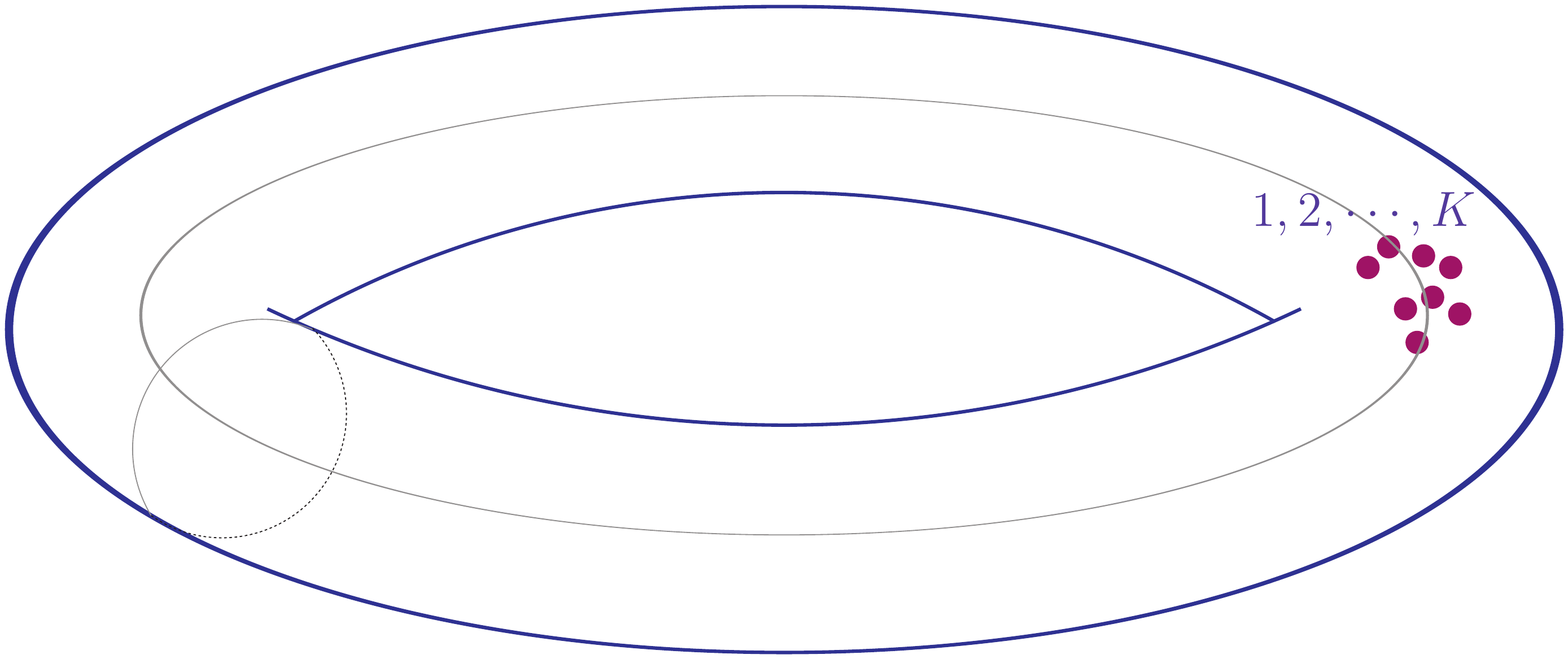}
\caption{\sl The strong coupling singular limit of the UV curve for the four-dimensional, ${\cal N}=2$ SCFT with ${\widehat A}_{K-1}$ quiver gauge group.}
\label{fig:circular-quiver-singular}
\end{figure}

As we discussed above, the singular limit we are interested in corresponds to bringing the $K$ punctures together, see figure \ref{fig:circular-quiver-singular}.
One nice feature of Gaiotto's description is that things that happen locally on the Riemann surface are independent of what happens in the rest of the Riemann surface. In particular, the process of two or more punctures colliding together is a local process, which knows nothing about the rest of the Riemann surface. It leads to the same physics whether it happens on a torus (as in our case), on a sphere (as in most cases that have been analyzed in the literature), or on any other Riemann surface, as long as there are enough moduli to bring the punctures together. When two minimal punctures come together, it has been argued that, for any $N$, the Riemann surface develops a long throat. One then obtains a weakly coupled, S-dual, $SU(2)$ gauge theory with a gauge coupling constant that vanishes as the punctures come together, coupled to a hypermultiplet in the doublet representation (corresponding to the sphere on the other end of the long throat) and to an additional SCFT that depends on what happens in the rest of the Riemann surface. 

\begin{figure}[tbp]
\vspace*{-2cm}
	\includegraphics[angle=-90,width=15cm]{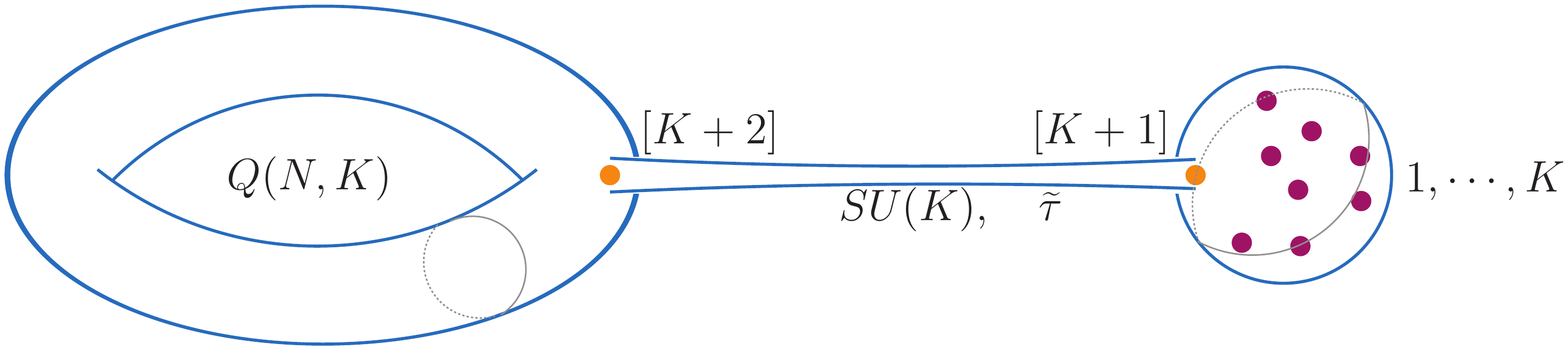}
\vspace*{-2cm}
\caption{\sl The  UV curve of the four-dimensional ${\cal N}=2$ superconformal quiver gauge theory in the singular limit, exhibiting
the separation into three decoupled sectors.}
\label{fig:circular-quiver-gaiotto1}
\end{figure}

Similarly, when $K$ punctures of this type come together, it has been argued \cite{Gaiotto:2009we,Gaiotto:2009hg,Chacaltana:2010ks,Chacaltana:2012zy} that, for large enough values of $N$, one obtains a weakly coupled $SU(K)$ gauge theory. In terms of the UV curve, the torus with $K$ coalescing punctures develops a long `throat' (see figure \ref{fig:circular-quiver-gaiotto1}), corresponding to a weakly coupled $SU(K)$ gauge group. On one side of this `throat' one has a torus with a single $SU(K)$ puncture 
(coupled to the $SU(K)$ gauge theory) and on the other side a sphere with $(K+1)$ punctures, the $K$ original punctures and an extra $SU(K)$ puncture
that is also coupled to the $SU(K)$ gauge theory. The coupling of the $SU(K)$ gauge theory goes to zero as the punctures come together and the `throat' becomes infinitely long, while the positions of the $K$ punctures on the sphere remain fixed (these depend on the relative distances between the punctures in the original picture, but not on their overall distance scale).

In this limit we have 3 separate theories, that are coupled only by the constraint of the $SU(K)$ gauge invariance:
\begin{itemize}
\item $Q(N,K)$, which is the four-dimensional ${\cal N}=2$ SCFT corresponding to the six-dimensional $A_{N-1}$ SCFT on a torus with a single $SU(K)$ puncture.
\item $P(K)$, which is the four-dimensional SCFT corresponding to a sphere with $K$ minimal punctures and one $SU(K)$ puncture. This theory is independent of $N$ (in the wrapped 5-brane picture, only some of the $N$ 5-branes wrap this part of the Riemann surface). 
\item The S-dual $SU(K)$ gauge theory, which gauges the diagonal $SU(K)$ group of the $SU(K)$ global symmetry of the $Q(N,K)$ theory and the $SU(K)$ global symmetry of the $P(K)$ theory.
\end{itemize}

The assignment of the different parameters and symmetries to the different components is the following. The parameter $\tau_0$ of the original four-dimensional ${\cal N}=2$ SCFT becomes a parameter of the $Q(N,K)$ theory, corresponding to
the complex structure of its torus. One additional parameter of the original theory becomes the coupling constant $\tilde \tau$ of the S-dual
$SU(K)$ gauge theory, while the other $(K-2)$ parameters become parameters of the $P(K)$ theory (recall that a sphere with $n$ punctures has
$(n-3)$ complex moduli). The $U(1)^{K-1}$ `relative' global symmetry of our quiver SCFT is a global symmetry of $P(K)$ -- it is not part of the S-dual $SU(K)$ gauge theory, in which $SU(K)$ is a gauge symmetry  rather than a global symmetry. 

\begin{figure}[tbp]
\vspace*{-2cm}	
\includegraphics[angle=-90,width=15cm]{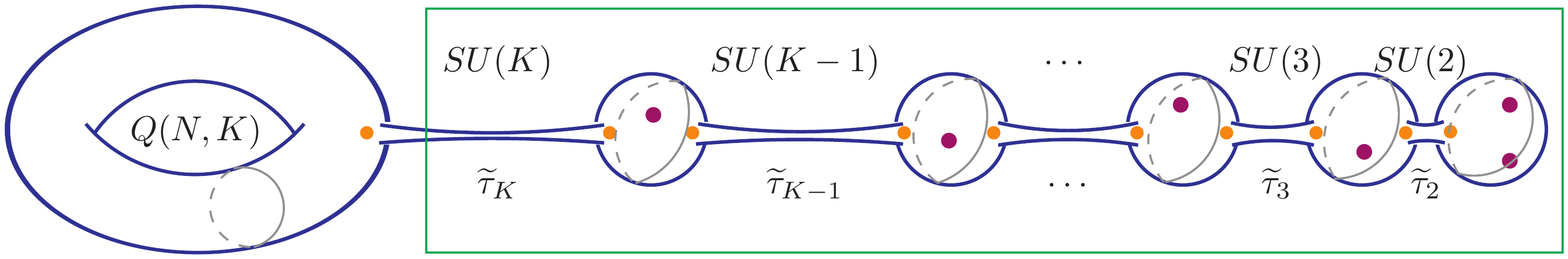}
\vspace*{-2cm}
\caption{\sl The weakly coupled S-dual, four-dimensional ${\cal N}=2$ superconformal quiver gauge theory.}
\label{fig:circular-quiver-gaiotto2}
\end{figure}

We can take the limit to the singular point in various different ways (for instance, by taking different ratios of
the $B_2$ fields on the 2-cycles as they go to zero). In all of these limits, we get the weakly coupled S-dual $SU(K)$ gauge theory, but we sit at different points in the parameter space of the $P(K)$ theory. In general, the $P(K)$ theory is strongly coupled, but there is a special region of its parameter space in which it is weakly coupled. In fact, this gives the easiest way to substantiate where the symmetries and parameters reside. We can
land in this region by first bringing together two punctures, then bringing together at a slower rate a third puncture, and so on, until all
punctures come together. Bringing together the first two punctures leads to a weakly coupled $SU(2)$ gauge group, bringing together
three punctures gives a weakly coupled $SU(3)$, and so on. Thus, in this limit the $P(K)$ theory becomes a weakly coupled gauge theory with gauge group\footnote{A very similar S-duality was recently discussed in \cite{McGrane:2014pma,Bergman:2014kza}, for $K$ punctures coming together on a sphere.}
\begin{equation}\label{new_gauge}
\hat{G} = SU(2)\times SU(3)\times \cdots \times SU(K-1),
\end{equation}
with a bifundamental hypermultiplet for each pair of adjacent gauge groups, another hypermultiplet in the doublet of $SU(2)$, and another
$K$ hypermultiplets in the fundamental representation of $SU(K-1)$ that couple as anti-fundamentals to the $SU(K)$ gauge theory described
above (see figure \ref{fig:circular-quiver-gaiotto2})\footnote{It is worth mentioning a parallel between this limit and gauge mediation. In the latter, a hidden sector comprising of a strongly interacting CFT is coupled through a weakly interacting messenger gauge theory to the weakly coupled visible sector. At leading order, the visible sector as well as the messenger gauge theory can be ignored, while the hidden sector is a CFT with a certain global symmetry. This global symmetry will be gauged weakly once the messenger interaction is turned on. In our case the gauge theory $\hat{G}$ serves the role of
the visible sector, the $SU(K)$ group is analogous to the messengers, and the $Q(N,K)$ theory to the hidden sector.}.
In this limit, the $(K-2)$  parameters of the $P(K)$ theory become the gauge couplings
of (\ref{new_gauge}). Note that the new gauge group has nothing to do with the
original $SU(N)^K$ gauge symmetry, but appears at strong coupling, in analogy with the S-dual gauge group in four-dimensional ${\cal N}=4$ super Yang-Mills theory, or with the $SU(2)$ gauge group in the strong coupling limit of the $SU(3)$ gauge theory with six flavors of fundamental hypermultiplets \cite{Argyres:2007cn}. This is of course possible because asking what gauge group a specific field theory has is an ill-defined question, except in the situation that its coupling is extremely weak. 

Since we have SCFTs for any value of the parameters, it is clear that the beta functions of the $SU(K)$ gauge group and of the gauge groups in
(\ref{new_gauge}) must all vanish. For the gauge groups in (\ref{new_gauge}), this is obvious from their matter content. For the S-dual $SU(K)$ gauge group, this comes from a cancelation between the contribution of its vector multiplets, the $P(K)$ theory (whose weakly coupled limit makes it clear that it contributes like $(K-1)$ fundamental hypermultiplets of $SU(K)$), and the $Q(N,K)$ theory
(which must, therefore, contribute like $(K+1)$ fundamental hypermultiplets of $SU(K)$).

Note that, in the S-dual description, the $U(1)^{K-1}$ `relative' $U(1)$ global symmetries of the original theory all act on 
the $P(K)$ factor. In its weakly coupled limit (\ref{new_gauge}), they map to the $U(1)^{K-1}$ global symmetries acting on the various
hypermultiplets described above.\footnote{It would be interesting to understand the precise mapping
between the `relative' $U(1)^{K-1}$ symmetries of the original $SU(N)^K$
theory, and those of the $P(K)$ theory.}
\section{The Singular Limit:  $AdS_5$ Bulk Perspective}
\label{section4}

In this section, we describe how to map our results of the previous section to the bulk and what our results imply about the ${\cal N}=(2,0)$ $A_{K-1}$ theory on $AdS_5\times \mathbb{S}^1$.

In the singular limit corresponding to the origin of the `moduli space', we expect some of the bulk physics to give the six-dimensional ${\cal N}=(2, 0)$ theory 
on $AdS_5\times \mathbb{S}^1$, while the rest of the bulk physics should decouple at low energies and should not depend strongly on where we are in the `moduli space'. As we will discuss in the next section, we can take a decoupling limit in which the six-dimensional $A_{K-1}$ theory completely decouples from the rest of the bulk physics. For our considerations in this section, however, the details of this limit will not be important. In our picture of the previous section, it is clear that in the singular limit the bulk physics which is not associated with the $A_{K-1}$ theory maps only to the $Q(N,K)$ sector of the four-dimensional SCFT, which does not depend on where we are in the `moduli space' \eqref{ads5modulispace} of the $A_{K-1}$ theory. The fields in the $A_{K-1}$ theory on $AdS_5\times \mathbb{S}^1$ thus map to the $P(K)$ sector, the $SU(K)$ gauge theory, and some part of the $Q(N,K)$ sector of the four-dimensional SCFT. In particular, the $(K-1)$ complex `moduli' of this theory map to the parameters of the $SU(K)\times P(K)$ theory.

As we saw above, in the singular limit, the four-dimensional ${\cal N}=2$ SCFT contains an S-dual $SU(K)$ gauge theory that becomes free and hence the SCFT acquires an infinite number of conserved {\sl higher-spin} currents~\footnote{Even more
conserved high-spin currents arise if we take a more specific limit, where some additional gauge groups in (\ref{new_gauge}) become free.}. 
The discussion above suggests that these currents should map by the AdS/CFT correspondence to the $A_{K-1}$ theory on $AdS_5 \times \mathbb{S}^1$. 
Naively, this implies that this bulk field theory must contain, at the origin of its `moduli space', massless high-spin gauge fields, including a massless graviton (corresponding
to the energy-momentum tensor of the $SU(K)$ gauge theory, which is separately conserved in this limit). This is very different
from what this theory behaves like in flat space, as we recalled above. Indeed, as we will review below, in flat space the
appearance of such massless higher spin gauge fields would be impossible, though it is not impossible on anti-de Sitter space.
Nevertheless, this seems like a very unlikely behavior for a local field theory. We will begin in the next subsection by suggesting a simpler alternative way to interpret the singular limit, in which there are no massless higher spin gauge fields on $AdS_5$. In subsection
\ref{implications}, we describe the implications of this alternative description and its relation to the behavior of the $A_{K-1}$ theory on $\mathbb{R}^{4,1}\times \mathbb{S}^1$.
In subsection \ref{highspin}, we return to the picture with massless high spin gauge fields and review why this picture is not impossible. In  subsection \ref{equivalence}, we suggest an interpretation for this picture that is consistent with our alternative description.

\subsection{A Simpler Description of the Physics on $AdS_5$}

Above we assumed that, following the rules of the AdS/CFT correspondence, the high-spin conserved currents arising from the weakly coupled S-dual $SU(K)$ gauge theory map to massless high-spin fields on $AdS_5$. However, there is also another
logical possibility, that this four-dimensional $SU(K)$ gauge theory lives on the boundary of $AdS_5$. This leads to a picture where the
$SU(K)$ and $P(K)$ sectors of the four-dimensional ${\cal N}=2$ SCFT in the singular limit live on the boundary of $AdS_5$, and
couple to the $Q(N,K)$ theory which is identified with the physics in the bulk of $AdS_5$ (including both the $A_{K-1}$ theory on $AdS_5\times \mathbb{S}^1$ and the rest of type IIB string theory on $AdS_5\times  \mathbb{S}^5/\mathbb{Z}_K$).

A complete classification of supersymmetry-preserving boundary conditions for the ${\cal N}=(2,0)$ ${A}_{K-1}$ theory on $AdS_5\times \mathbb{S}^1$ is not known. For our description, we only need to assume that boundary conditions exist that allow for a coupling to this four-dimensional ${\cal N}=2$ theory living on the boundary of $AdS_5$. This is somewhat similar to the boundary conditions of the ${\cal N}=4$ SYM theory on $AdS_4$, which allow for a coupling
to general three-dimensional ${\cal N}=4$ SCFTs living on the boundary \cite{Gaiotto:2008ak,Gaiotto:2008sa}. However, note that in our case the
theory on the boundary is not precisely conformal by itself, but rather the $SU(K)$ gauge theory has a non-zero beta function (even after the coupling to $P(K)$) which must be precisely canceled by its coupling to the bulk fields. This means that the bulk physics must contain a strongly coupled five-dimensional $SU(K)$ gauge theory which couples to the four-dimensional $SU(K)$ gauge theory on the boundary (such that the four-dimensional gauge fields are identified with the boundary values of the five-dimensional gauge fields). But this implies that we must have an $SU(K)$ gauge theory living on $AdS_5$ {\it everywhere} in the `moduli space' \eqref{ads5modulispace}, since in the four-dimensional SCFT the $SU(K)$ gauge symmetry is unbroken for any value of the exactly marginal couplings. We are thus led to the surprising conclusion that the six-dimensional ${\cal N}=(2,0)$ $A_{K-1}$ theory on $AdS_5 \times \mathbb{S}^1$ includes, everywhere on its `moduli space' \eqref{ads5modulispace}, a five-dimensional $SU(K)$ supersymmetric gauge theory on $AdS_5$, coupled to a specific four-dimensional ${\cal N}=2$ theory on the boundary of $AdS_5$. We will discuss the relation of this $SU(K)$ theory to the one that appears at the origin of the moduli space \eqref{5dmodulispace} of the $A_{K-1}$ theory on $\mathbb{R}^{4,1}\times \mathbb{S}^1$ below.

While this interpretation gets rid of the problem of having massless higher spin gauge fields in the bulk, it immediately leads to two
other problems.

\subsubsection{First Problem : Continuity of the Description}
\label{continuity}

The first problem is that this description is very different from our description of the physics far from the singular point
(and in particular at the orbifold point). The physics of Type IIB string theory on $AdS_5\times \mathbb{S}^5/\mathbb{Z}_K$ 
changes continuously as we vary the parameters, including the periods of the two-forms on vanishing 2-cycles, since the 
dual four-dimensional ${\cal N}=2$ SCFTs vary continuously as the exactly marginal coupling constants are varied continuously.
However, far from the singular point of the `moduli space' we saw that we have a $U(1)^{K-1}$ gauge theory in $AdS_5$, and no fields living on the
boundary, while we are now suggesting that near (but not necessarily at) the singular point we have an $SU(K)$ gauge theory
in $AdS_5$, and an extra four-dimensional ${\cal N}=2$ theory living on the boundary.

As we approach the singular point, the five-dimensional $U(1)^{K-1}$ gauge theory becomes strongly coupled due to the presence of light charged fields,
and (as we will show below) the five-dimensional $SU(K)$ gauge theory is always strongly coupled, so there is no sharp contradiction in having both
descriptions be valid (and perhaps related by some strong-weak coupling duality). Moreover, when the theory on $AdS_5$ is not weakly coupled,
there is no sharp distinction between the fields living in the $AdS_5$ bulk and on the boundary. Indeed, there are known examples where S-duality changes the degrees of freedom on the boundary \cite{Gaiotto:2008ak,Gaiotto:2008sa}, and even relates a boundary condition with no extra degrees of freedom to one with extra degrees of freedom on the boundary. This suggests that asking about the existence of extra degrees of freedom at the boundary is not a well-posed question, and the two descriptions could be equivalent.

Note that even though in our suggested description we have an $SU(K)$ gauge symmetry in the bulk, this gauge group
does not break far away on the `moduli space' into the $U(1)^{K-1}$ gauge group that we have there (as was the
case for the ${\cal N}=(2,0)$ $A_{K-1}$ theory on $\mathbb{R}^{4,1}\times \mathbb{S}^1$). Indeed, the four-dimensional $U(1)^{K-1}$ global symmetry, that is related to
this $U(1)^{K-1}$ gauge group, is present everywhere on the `moduli space' and is never enhanced; note that in
our description of the physics near the singular point, this global symmetry acts only on the $P(K)$ theory living on the boundary of $AdS_5$. So the five-dimensional $SU(K)$ gauge fields in our description must have a different origin than the $U(1)^{K-1}$ gauge fields.
As we approach the singular point in type IIB string theory,
the D3-branes wrapped on the vanishing 2-cycles become light, and there must be some complete rearrangement of the degrees
of freedom, which leads to the new fields living on the boundary, and to the new $SU(K)$ gauge symmetry
in $AdS_5$. Note that this gauge symmetry does not give rise to a global symmetry in the four-dimensional
${\cal N}=2$ SCFT, since the global currents that it leads to are gauged by their coupling to the four-dimensional $SU(K)$ gauge theory on the boundary of $AdS_5$.
In our scenario the $U(1)^{K-1}$ global symmetry of the four-dimensional ${\cal N}=2$ SCFTs continuously interpolates
between being associated with the $U(1)^{K-1}$ massless gauge fields in the bulk when we are far
from the singular point, and being associated with the $P(K)$ theory on the boundary when we are close to the singular point.
The relation between the two pictures is presumably some novel form of strong-weak coupling duality.

\subsubsection{Second Problem : Boundary Condition for the Bulk Gauge Fields}

A second problem is that it is usually stated in the AdS/CFT correspondence that there is only a single allowed boundary condition for five-dimensional gauge fields on $AdS_5$ \cite{Marolf:2006nd}, which is the one leading to a global symmetry in the
dual four-dimensional CFT. Here we claim that we can choose another boundary condition for our five-dimensional $SU(K)$ gauge theory on $AdS_5$, which does not lead to a global symmetry, but rather couples it to a four-dimensional $SU(K)$ gauge theory living
on the boundary of $AdS_5$. How is this consistent with the standard lore ?

From the four-dimensional field theory point of view, whenever we have a non-anomalous global symmetry $H$ (with no global
anomalies), we are allowed to 
couple it to four-dimensional gauge fields with gauge group $H$. This should be possible, in particular, for global symmetries $H$ related by the AdS/CFT correspondence to five-dimensional gauge fields (if these five-dimensional gauge fields do not have Chern-Simons terms). The standard rules of the AdS/CFT correspondence imply that one way to
describe this process is to add four-dimensional gauge fields for $H$ on the boundary of $AdS_5$ (with or without a four-dimensional kinetic term), and to identify them with the
boundary value of the five-dimensional $H$ gauge fields in the bulk. This is because, in the standard choice of boundary conditions, these 
boundary values are identified with the couplings to the global symmetry currents of the global symmetry $H$.

So why is it usually claimed that such modified boundary conditions are not allowed on $AdS_5$ ? The point is that this claim is made in the
context of boundary conditions preserving the four-dimensional conformal symmetry (and the corresponding isometries of $AdS_5$). Starting from a theory with four-dimensional conformal symmetry, the process described above only preserves this symmetry if when we gauge $H$, its
beta function exactly vanishes. In particular, the contribution from the five-dimensional
gauge theory to the beta function of the four-dimensional gauge theory $H$ must precisely cancel all other contributions.

When we have a weakly coupled gauge theory $H$ on $AdS_5$, with a gauge coupling $g_H^{(5)}$ satisfying $(g_{H}^{(5)})^2 \ll R_{\rm AdS}$, 
it is easy to compute its contribution to the four-dimensional beta
function that would arise if we couple it to four-dimensional gauge fields. This contribution is proportional to the two-point
function of the global symmetry currents, which goes as $R_{\rm AdS} / (g_H^{(5)})^2 \gg 1$ in this limit. Since in a unitary field theory this
contribution is positive, and since the only negative contribution to the beta function comes from the four-dimensional gauge fields of
$H$ itself, there is no way to obtain a vanishing four-dimensional beta function in this case. The usual statement that only one
boundary condition is allowed assumes a weakly coupled gauge theory on $AdS_5$, and also that the conformal symmetry is preserved. In this context, this statement is indeed correct.

However, our discussion above implies that when the five-dimensional gauge coupling is large enough so that it gives a small contribution to the four-dimensional beta function, it may be possible to gauge the corresponding global symmetry
while still preserving the conformal symmetry. In particular, we claim that this is precisely what happens in
our scenario for describing the physics near the singular point of the $\mathbb{S}^5/\mathbb{Z}_K$ orbifold.
Indeed, our computation in the previous section implies that the five-dimensional bulk physics should contribute to the four-dimensional
beta function of $SU(K)$ exactly the same as $(K+1)$ hypermultiplets in the fundamental representation.
If we translate this to the five-dimensional gauge coupling, we find that 
\begin{equation}
(g_{SU(K)}^{(5)})^2 \quad \simeq \quad R_{\rm AdS} / (K+1), 
\end{equation}
implying that the five-dimensional $SU(K)$ gauge theory is strongly coupled\footnote{Naively it is weakly coupled for large $K$, but
its 't Hooft coupling still remains of order one at the AdS scale.}. Of course this computation is not really reliable when
the five-dimensional gauge theory is strongly coupled. But in any case we conclude that the five-dimensional gauge theory must be strongly coupled, thus helping also to avoid the first problem mentioned above. This is not surprising since
we expect that, like in the ${\cal N}=(2,0)$ theory in flat space, $(g_{SU(K)}^{(5)})^2$ should be of order $R_{{\mathbb S}^1} = R_{\rm AdS}$. Presumably,
the cancelation of the beta function only happens for a specific ratio of $R_{{\mathbb S}^1}/R_{\rm AdS}$, which is precisely the one we get in the type IIB string theory.



\subsection{Implications}
\label{implications}

In the previous subsection, we suggested a scenario for what happens near the singular point in the
`moduli space', and described how this scenario gets around two apparent problems. In this subsection
we discuss several surprising implications of this scenario.

\subsubsection{Other Choices of Boundary Conditions and the Holographic Dual of $Q(N,K)$}
\label{otherbc}

Our discussion of the second problem has an interesting implication that will help us to understand the relation between the five-dimensional $SU(K)$ gauge theory that we found and the one that appears on $\mathbb{R}^{4,1}\times \mathbb{S}^1$. 
We argued that we can give
alternative boundary conditions to the five-dimensional $SU(K)$ gauge theory on $AdS_5$, and that these alternative
boundary conditions naturally arise from type IIB string theory in the scenario we discussed. However,
we can also ask what happens if we give standard boundary conditions to the $SU(K)$ gauge fields appearing in type IIB string theory on $AdS_5\times \mathbb{S}^5/\mathbb{Z}_K$,
such that they correspond to an $SU(K)$ global symmetry. Our discussion above implies that this
other choice of boundary condition, in which we do not need to add any extra fields on the boundary, 
should correspond to a dual four-dimensional SCFT which is precisely the $Q(N,K)$ theory. At the very least, we can say that the low-energy limit of the holographic dual to $Q(N,K)$ should be the same as that of type IIB string theory on $AdS_5\times \mathbb{S}^5/\mathbb{Z}_K$ at its singular orbifold point, but with these standard
boundary conditions for the $SU(K)$ gauge fields.

The $Q(N,K)$ theory has only a single coupling constant, which is identified with the dilaton/axion
in the bulk. This implies that, with these other boundary conditions, the ${\cal N}=(2,0)$ $A_{K-1}$ theory living at the orbifold point of $\mathbb{S}^5/\mathbb{Z}_K$ has
no `moduli space', viz. it always sits at the origin. Recall that in any case, in our picture that included the $SU(K)$ gauge theory on $AdS_5$, the motion on the `moduli space' was not described by
turning on five-dimensional scalar fields of this gauge theory, but rather by changing the coupling constants of the four-dimensional $SU(K)\times P(K)$ theory living on the boundary of $AdS_5$. In this sense, our picture is consistent. Presumably, for either boundary condition, the $SU(K)$-adjoint scalar fields on $AdS_5$ have some potential which does not allow them to obtain expectation values.

Note that the ${\cal N}=(2,0)$ $A_{K-1}$ theory, in which this standard choice of boundary conditions
should certainly be possible, involves only a sector of the $Q(N,K)$ theory (as discussed in more
detail in the next section). But
it is interesting to speculate that perhaps there is some type IIB string theory that realizes
these standard boundary conditions for the five-dimensional $SU(K)$ gauge fields, and that is dual to the full $Q(N,K)$ theory. In such a string theory the $\mathbb{S}^5/\mathbb{Z}_K$ orbifold would always be frozen to its singular point; since the
physics at this point is not weakly coupled, it is not clear how to test if such a string theory exists or not. Recall that with orientifolds, it is known that a $\mathbb{C}^2/\mathbb{Z}_2$ singularity can 
be frozen \cite{Polchinski:1996ry}, so perhaps an analogous mechanism exists also in our case.

We mentioned above that the boundary conditions for which the five-dimensional $SU(K)$ gauge theory is coupled to boundary gauge fields (coupled to $P(K)$) are only possible for a specific ratio of the radii of $AdS_5$ and $\mathbb{S}^1$,
since only for this specific ratio its contribution to the beta function of the four-dimensional $SU(K)$ gauge theory is the right one for preserving the four-dimensional conformal symmetry. In the six-dimensional ${\cal N}=(2,0)$ $A_{K-1}$ theory on $AdS_5\times \mathbb{S}^1$, it seems that the standard boundary conditions for which the $SU(K)$ group is a global symmetry should be possible and preserve supersymmetry for any ratio of radii \footnote{For equal radii the fact that supersymmetry is preserved follows from the conformal transformation to flat space discussed in appendix \ref{appendixA}, but supersymmetry is actually preserved for an arbitrary ratio of radii, as for $\mathbb{S}^d \times \mathbb{S}^1$ \cite{Talya}.}. However, presumably there is no string theory embedding for the theory with these boundary conditions at a generic ratio of radii, because there is no additional
exactly marginal deformation of the $Q(N,K)$ theory that could correspond to it. Thus, it seems that for a generic ratio we
cannot map the six-dimensional ${\cal N}=(2,0)$ $A_{K-1}$ theory on $AdS_5\times \mathbb{S}^1$ to a sector in some four-dimensional SCFT.

In particular, in the six-dimensional theory with the standard boundary conditions, we can take the limit where the radius of $AdS_5$ is much larger than the radius of
$\mathbb{S}^1$, and in this limit we should approach the theory on $\mathbb{R}^{4,1}\times \mathbb{S}^1$.
It is natural to expect that the five-dimensional $SU(K)$ gauge group that we found on $AdS_5$ will become in this limit the five-dimensional $SU(K)$ gauge group on $\mathbb{R}^{4,1}$; note that this limit involves the standard
boundary conditions, for which this $SU(K)$ is always unbroken. From this point of view, it is not
surprising that we found a five-dimensional $SU(K)$ gauge group also in the theory on $AdS_5\times \mathbb{S}^1$.
Note that since to take this
limit we have to use the standard boundary conditions which have no `moduli space', there is
no direct connection between the moduli space \eqref{5dmodulispace} on $\mathbb{R}^{4,1}\times \mathbb{S}^1$,
on which the $SU(K)$ gauge group is spontaneously broken, and the `moduli space' \eqref{ads5modulispace}
of the theory with alternative boundary conditions on which it is not, thus avoiding a potential contradiction. 

One might have thought that the two moduli spaces would be related, since in type IIB string theory
both involve fields from the twisted sector of the $\mathbb{Z}_K$ orbifold, and in both cases D3-branes
wrapped on the 2-cycles of the orbifold times the circle become light near the origin of the moduli space. In particular,
since in flat space these wrapped D3-branes (wrapped on 2-cycles times the $\mathbb{S}^1$) 
give the W-bosons of the $SU(K)$ gauge theory, one might have thought that they have the same role 
also on $AdS_5\times \mathbb{S}^1$. However, our analysis implies that this cannot possibly be correct and that
the $SU(K)$ gauge fields on $AdS_5$ do not come from wrapped D3-branes (since the latter
do become massive on the `moduli space'). Apparently, there is some intricate reorganization of the degrees of freedom involved in the interpolation from $AdS_5$ to flat space. As we discussed, there is no string theory picture of the intermediate steps involved in this interpolation. Therefore, we
cannot ask whether the $SU(K)$ comes from wrapped D3-branes or not for generic values of
the ratio of radii.

\subsubsection{The Metric on the Moduli Space}

In flat space, the metric on the moduli space (\ref{5dmodulispace}) of the ${\cal N}=(2,0)$ theory is
flat, and supersymmetry prevents it from receiving any quantum corrections. In particular, the origin
of the moduli space is at a finite distance. In the AdS/CFT correspondence, the metric on the `moduli
space' in $AdS_5$ is mapped to the Zamolodchikov metric, which is proportional to the 2-point
functions for the exactly marginal deformations in the dual four-dimensional SCFT. 

\begin{figure}[tbp]
\centering
\vspace*{-2cm}	
\includegraphics[angle=0,width=12cm]{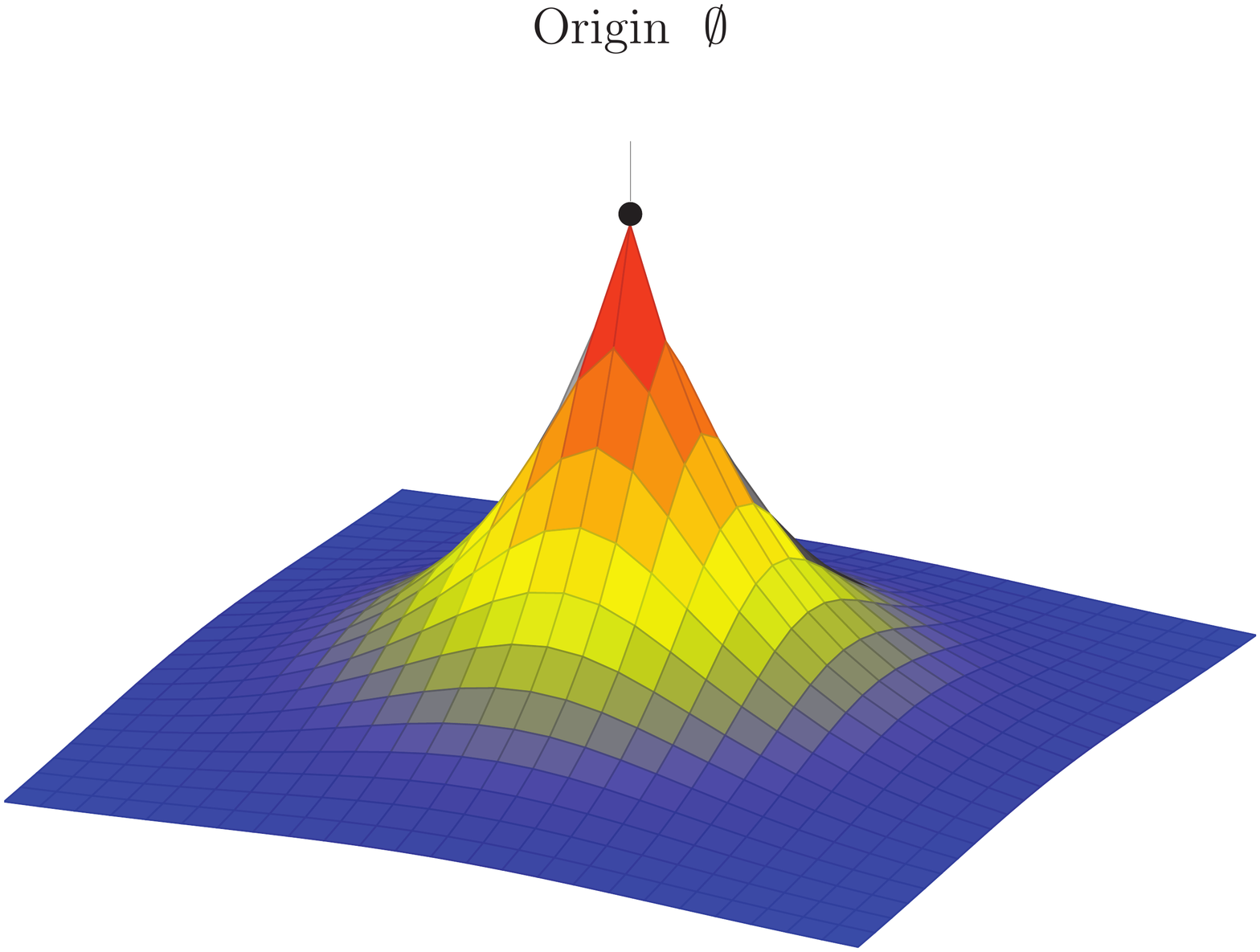}
\vspace*{0cm}
\caption{\sl The `moduli space' of the six-dimensional $(2,0)$ theory on $AdS_5 \times \mathbb{S}^1$. The shape highlights the geometry of the moduli space, where the point $0$ refers to the origin, sitting at infinite distance away from an interior point. The coloring highlights different effective descriptions of the same theory at different values of the theory's parameters. }
\label{modspacefig}
\end{figure}

In our case, we saw that at the origin of the `moduli space' (\ref{ads5modulispace}), the coupling constant of the four-dimensional 
S-dual $SU(K)$ gauge theory goes to zero. However, the Zamolodchikov metric for a weakly coupled gauge
theory goes as:
%
\be
\rmd s^2 \simeq { {|d{\widetilde \tau} |^2} \over ({\rm Im} (\widetilde{\tau}))^2},
\ee
so this implies that the origin of the `moduli space' on $AdS_5$ (which we mapped to $\widetilde{\tau} \to i\infty$) is
infinitely far away, see figure \ref{modspacefig} \footnote{The Zamolodchikov metric for the quiver gauge theory can
actually be computed exactly, for any $N$ and any coupling constants, using localization on $\mathbb{S}^4$ \cite{Pestun:2007rz,Gerchkovitz:2014gta,Gomis:2014woa}, as can
extremal correlation functions.}.

Furthermore, our discussion implies that the origin of the `moduli space' is not just a point but a 
$(K-2)$-dimensional space, corresponding to the parameter space of the $P(K)$ theory. This is most 
cleanly seen from the Gaiotto picture, which shows that the space of coupling constants of the $P(K)$ theory 
is the moduli space of a $(K+1)$-point punctured sphere $\mathbb{S}^2_{K+1}$.

Thus, we argue that there are large quantum corrections to the metric on the `moduli space'
(\ref{ads5modulispace}), such that near the origin this `moduli space' develops a semi-infinite
`throat', and that fibered over this `throat' we have a $(K-2)$ complex-dimensional space
which is the space of parameters of the $P(K)$ theory. This space itself has extra `throats',
corresponding to the extra weakly coupled limits of the gauge groups in (\ref{new_gauge}).

Note, however, that in the picture of the theory on $AdS_5$ that we described in the previous subsection, the exactly marginal
deformations discussed here do not correspond to bulk fields, but to coupling constants of the theory
on the boundary of $AdS_5$. Thus, in this alternative description, we do not really have
a `moduli space' in the bulk. The comments above refer to the original description of the
physics in $AdS_5$, in terms of scalar fields in $U(1)^{K-1}$ vector multiplets parameterizing
a `moduli space'; in this description we see that large corrections to the metric must arise
near the origin.

\subsection{Higher-Spin Gauge Theory Description}
\label{highspin}

Let us now return to the other picture of the physics on $AdS_5$, in which we do not have any fields
living on the boundary of $AdS_5$. This picture is more directly related to the physics far on the
`moduli space'. However, since in the dual SCFT we have a free four-dimensional $SU(K)$ gauge theory at the origin of the `moduli space', with higher spin conserved currents, this picture necessarily has at the origin of the `moduli 
space' massless higher-spin fields (including an extra massless graviton) on $AdS_5$. We argued above
that the four-dimensional $SU(K)$ gauge theory is associated with the $A_{K-1}$ ${\cal N}=(2,0)$ theory on $AdS_5\times \mathbb{S}^1$ in the bulk,
so these massless higher-spin fields must arise from this local field theory.

In flat space, it is well-known (see, for instance, \cite{Weinberg:1980kq}) that a local Poincar\'e invariant quantum field theory cannot give rise at low energies to a massless graviton, or to massless higher-spin fields. In anti-de Sitter and de Sitter spaces, no analogous theorem is known. This is related to the absence of a van Dam-Vainshtein-Zakharov discontinuity: the number of degrees of freedom in these spaces does not change discontinuously as the mass of a graviton or of higher-spin fields goes to zero, but rather the massive spin-$s$ field continuously goes into a massless spin-$s$ field together with lower-spin fields (by a generalized Higgs mechanism). So this picture does not lead to any immediate contradictions.

Note that the AdS/CFT correspondence does not imply that these massless high-spin fields should be weakly coupled in $AdS_5$. In particular, the four-dimensional S-dual gauge theory that arises at the singular point is independent of $N$, so its bulk dual description should involve strongly coupled fields, whose interactions are not suppressed by Newton's gravitational constant $G_5$ which is proportional to $1/N^2$. What the AdS/CFT correspondence does imply is that there must exist states on $AdS_5$ which carry the same quantum numbers as the massless graviton and higher spin fields described above. These states should be exactly massless when the periods of the two-form potentials vanish so that we are at the origin of the `moduli space' \eqref{ads5modulispace}, and should continuously obtain a mass by a generalized Higgs mechanism when we move away from the origin. Note that this is consistent with the representations of the four-dimensional ${\cal N}=2$ superconformal algebra: multiplets containing energy-momentum tensors and higher-spin fields can continuously become conserved when varying parameters or coupling constants.

Where can these extra massless states come from in the bulk? It is natural to view the extra light states as arising from the D3-branes wrapped on the two-cycles, 
which become light as we go to the origin of the `moduli space'. We argued above that the particles arising from these D3-branes that are also
wrapped on $\mathbb{S}^1$ are not BPS and do not become massless at the origin. However, the strings which are not wrapped on this circle are BPS and do become
tensionless at the origin. It is plausible that these `tensionless strings' are related to the massless high-spin fields (even though they do not give rise to such
fields at the origin of moduli space in $\mathbb{R}^{4,1}\times \mathbb{S}^1$).


\subsection{Equivalence Between the Two Descriptions}
\label{equivalence}

In the previous subsections, we saw two different descriptions of the physics near the origin of the
`moduli space'. In one description we argued that we should have (everywhere on the `moduli space') new fields living on the boundary of $AdS_5$,
and a five-dimensional $SU(K)$ gauge theory in the bulk. In the second description, we have at the origin of the `moduli space' massless high-spin fields. The first description has the advantages that it includes the $SU(K)$ gauge theory on $AdS_5$ that we expect to find (albeit with a different origin for the W-bosons), that it does not
contain massless high-spin fields, and that by a small change in the boundary conditions it gives a dual for
the $Q(N,K)$ theory. The second description has the advantage that it is more closely related to the 
weakly coupled description of the physics far out
on the `moduli space', which does not involve any fields on the boundary.

We would like to argue that the two descriptions are two different points of view on the same
physics.
%
In the first description, we have a four-dimensional $SU(K)\times P(K)$ theory living on the boundary of $AdS_5$.
But, this four-dimensional gauge theory should also have an AdS/CFT dual, which contains massless high-spin
fields (that are not weakly coupled for finite $K$) when the $SU(K)$ gauge theory becomes free. It is 
plausible that replacing this theory by its five-dimensional holographic dual gives precisely our second description.
A priori, if we just take some boundary theory coupled to a bulk
theory, and replace the boundary theory by its AdS dual, its coupling to the bulk theory would 
not be local on AdS space. However it is possible that in our case, even though we do not know how to directly analyze this, we do get local couplings
on $AdS_5$.

Another possibility, which is not obviously different, is that the two descriptions are related
by some strong-weak coupling duality; as mentioned above there are known cases where
such dualities relate theories with extra fields on the boundary, to other theories without
such fields.

In any case, the bottom line of our analysis is that there is some sector of the four-dimensional ${\cal N}=2$ SCFT which captures the physics of the $A_{K-1}$ ${\cal N}=(2,0)$ theory on $AdS_5\times {\mathbb{S}}^1$.
We will discuss in the next section precisely in which sense such a sector can describe a local field
theory on $AdS_5$. We provided two possible descriptions of the physics of the ${\cal N}=(2,0)$
theory on $AdS_5$, that are presumably equivalent, and that illustrate different aspects of
this physics. The two descriptions have different gauge symmetries on $AdS_5$, but  gauge symmetries are not physical -- the only well-defined
question is what are the gauge-invariant states, or equivalently, the gauge-invariant operators in the dual four-dimensional SCFT. 
Evidently, these are the same in both descriptions.

\section{Rigid Holography from Decoupling Limits in Anti-de Sitter Space}
\label{reverse}

In the previous sections we were somewhat imprecise about exactly which sector of the four-dimensional ${\cal N}=2$ SCFT of $\widehat{A}_{K-1}$ type is related to the six-dimensional ${\cal N} = (2,0)$ $A_{K-1}$ theory on $AdS_5 \times \mathbb{S}^1$ and how. In this section, we will make this more precise. Actually, the issue at hand is more general, since there are many other field theories on $AdS$ space that can be realized as low-energy limits of gravitational theories on $AdS$, and the latter have CFT duals. Since we could not find a systematic discussion of this issue in the literature, we shall spend some time explaining this `rigid holography' in broader and general contexts, and briefly describe some other examples.

\subsection{General Considerations}

In a variety of situations in the AdS/CFT correspondence, we have branes or defects in the bulk that wrap an $AdS_p\times M_q$ subspace of the full $AdS_{d+1} \times M_{D-d}$ background (where $D=9$ for string theory, or $D=10$ for M theory)\footnote{Our statements can easily be generalized also to branes and defects that wrap asymptotically AdS spaces, and also other types of holographic spaces, but we will not discuss them here. See \cite{nakayamarey} for a specific class of such situations.}. In this case, the $AdS_{d+1}$ side of the correspondence contains both bulk modes and brane/defect modes. Both types of modes are dual to operators in the dual $d$-dimensional CFT. For $p < d+1$, the brane/defect modes are dual to operators living on a $(p-1)$-dimensional defect in the CFT, while for $p=d+1$, they are dual to a subset of the local operators of the full $d$-dimensional CFT.

In a generic situation, the separation in $AdS_{d+1}$ between brane/defect modes and other modes is not sharp. The brane/defect interacts with the bulk, and brane/defect modes mix with bulk modes. However, in some cases, we can take a decoupling limit in the bulk theory on $AdS_{d+1}$ that decouples the theory on the brane/defect from the rest of the bulk physics. In flat space, we can often decouple branes/defects by looking at low-energy limits, $E \ll M_{\rm pl}$ and $E \ll M_{\rm st}$. We can do the same also in $AdS$ space, and obtain decoupling limits that are related to field theories living on the $AdS$ space. In this case, we need to take the decoupling limit while keeping $E R_{\rm AdS}$ finite and fixed, namely,  we need to take a similar limit to the one we take in flat space, but with the dimensionless parameter $R_{\rm AdS} M_{\rm pl} \gg 1$. Moreover, if we want to remain with a local field theory on the branes, we also need to take $R_{\rm AdS} M_{\rm st} \gg 1$. Typically, this regime corresponds to an infinite $N$ limit from the viewpoint of the dual $d$-dimensional CFT, but the number of branes $K$ remains finite so that a specific field theory is decoupled on $AdS_p\times M_q$. 
%
%

In the situation that such a decoupling limit can be taken, the decoupling implies that correlation functions between the CFT operators that are dual to the brane/defect modes and the CFT operators that are dual to other modes vanish (in the convention where we normalize all 2-point functions to one):
\be
\left\langle {\cal O}_{\rm bulk} \cdots {\cal O}_{\rm brane} \cdots \right\rangle \rightarrow 0.
\ee
Since we are taking the Planck scale to infinity, the $n$-point correlators of `single-trace' bulk modes with themselves also typically go to zero for $n>2$,
\be
\left\langle {\cal O}_{\rm bulk} {\cal O}_{\rm bulk} {\cal O}_{\rm bulk} \cdots \right\rangle \rightarrow 0.
\ee
The excitations localized on the brane/defect thus map in this limit to a sector of the CFT that is closed under the operator product expansion. This sector is `dual' to the low-energy theory living on the brane/defect, which is some field theory on $AdS_p\times M_q$. There is an important novelty, though. The correlation functions of the brane/defect modes in the bulk in general do not define a decoupled CFT by themselves, since the brane/defect modes do not usually include a massless spin-2 graviton. So, in most cases the dual sector of the CFT would not have a conserved energy-momentum tensor, which is one of the standard requirements from a CFT \footnote{However, the decoupled sector does have conserved energy and momentum charges, at leading order in the bulk gravitational interactions that are sent to zero. This does not imply the existence of a local energy-momentum tensor, since this sector does not have to be a local field theory.}.

In some cases, in the decoupling limit, we get a non-interacting theory living on the branes/defects, and the decoupled sector has non-zero two-point correlation functions only. It is then a generalized free field theory on the CFT side. In other cases, the theory on the branes/defects can be non-trivial, and this decoupled sector can have higher correlation functions among brane/defect modes that do not vanish in the decoupling limit,
\be
\left\langle {\cal O}_{\rm brane} {\cal O}_{\rm brane} {\cal O}_{\rm brane} \cdots \right\rangle \rightarrow {\rm finite}.
\ee
 The higher-point correlation functions in such cases would not be suppressed by powers of $N$, as is the case for the correlation functions among the bulk operators, and the operators ${\cal O}_{\rm brane}$ may be identified in the $d$-dimensional CFT by the fact that they are not generalized free fields. In these cases, the decoupled sector is strongly coupled compared to the bulk sector.
%
%
Note that, in particular, the operator product expansion of gauge-invariant operators in the brane/defect sector to the total energy-momentum tensor is suppressed:
\be
{\cal O}_{\rm brane} {\cal O}_{\rm brane} \simeq O\left(\frac{1}{N} \right) T_{mn} + \cdots.
\ee
%

In the decoupling limit, the physics on the gravitational side includes some decoupled $(p+q)$-dimensional field theory on $AdS_p\times M_q$ and, by construction, this decoupled theory includes the non-zero correlation functions of the $(p-1)$-dimensional operators corresponding to brane/defect modes, discussed above. These correlation functions give the response of this field theory to local sources at the boundary of $AdS_p$. Thus, we can learn about this response by rigid holography from the `dual' CFT. Of course, this bulk field theory is in many cases a local field theory put on a curved background so it also contains many more observables, namely, local operators located anywhere in $AdS_p\times M_q$, or equivalently, sources put anywhere in this space. However, these other observables do not have any obvious relation to the decoupled sector of the `dual' CFT.
If the field theory on $AdS_p\times M_q$ is conformal, these two classes of bulk observables can be distinguished: the bulk observables defined at asymptotic infinity are not deformations of the bulk observables defined at finite interior separation.  

 In general, to specify a field theory on $AdS$ space, one must also specify boundary conditions (that may be constrained by various (super)symmetries). The constructions above give us specific boundary conditions at the boundary of $AdS_p$ for the decoupled field theory on the branes/defect. We leave it as an interesting program for the future to apply the AdS/CFT correspondence for studying and possibly classifying general boundary conditions for field theories on $AdS_p\times M_q$.

\subsection{Examples}

Let us illustrate this general consideration with several familiar examples that involve branes in the $AdS_5\times \mathbb{S}^5$ background of Type IIB string theory, dual to the four-dimensional ${\cal N}=4$ supersymmetric Yang-Mills theory with gauge group $SU(N)$.

\subsubsection{D5-branes}

As our first example, consider $K$ overlapping D5-branes wrapping $AdS_4\times S^2$ \cite{Karch:2000ct,Karch:2001cw,Karch:2000gx}. This arises from $K$ D5-branes intersecting $N$ D3-branes along $\mathbb{R}^{2,1}$, in the near-horizon limit of the latter branes. So, in the field theory, it corresponds to having $K$ hypermultiplets in the fundamental representation of $SU(N)$ living on a codimension-one defect in the $(3+1)$-dimensional spacetime of the ${\cal N}=4$ SYM theory. Clearly, in general, the defect modes couple to the rest of the four-dimensional CFT. This
configuration preserves a three-dimensional ${\cal N}=4$ superconformal symmetry \cite{DeWolfe:2001pq},
that can be used to classify the operators and to constrain their correlation functions.

The D5-brane worldvolume dynamics is described at low energies (below the string scale) by 
a six-dimensional $U(K)$ gauge theory living on $AdS_4\times \mathbb{S}^2$. The gauge coupling of this theory in units of the AdS radius (which sets the typical scale) is
\be
{R_{\rm AdS}^2 \over g_{\rm YM}^2} \simeq {R_{\rm AdS}^2 \over  {g_{\rm st} \ell_{\rm st}^2}} \simeq \sqrt{N \over g_{\rm st}}.
\ee
As mentioned above, in order to decouple this theory from the bulk, we need to take $R_{\rm AdS} M_{\rm pl} \simeq N^{1/4}$ to infinity. We
see that in this limit the gauge coupling on the D5-branes vanishes, unless
we also scale $g_{\rm st}$ with $N$. Of course, if we do that, we would need to perform a
Type IIB S-duality transformation in the bulk for a better description; this S-duality turns
the D5-branes into NS5-branes, that we will discuss in the next subsection.

In the present situation, we can discuss two different decoupling limits. One option is to take
large $N$ with $g_{\rm st}$ fixed. In this case, the theory on the D5-branes becomes free. Moreover, the mass of all the massive open string states goes to infinity (in units of the AdS radius), so the theory on the D5-branes becomes simply the six-dimensional, free $U(K)$ gauge theory on $AdS_4\times \mathbb{S}^2$.
This theory is `dual', in the sense discussed above, to a sector of the theory living on the defects
in the four-dimensional CFT, in the same limit of large $N$ and fixed ${\cal N}=4$
super Yang-Mills coupling. We see that in this limit the only surviving modes are a single Kaluza-Klein tower of chiral operators, whose lowest component is the $U(K)$ supercurrent multiplet of the three-dimensional ${\cal N}=4$ supersymmetry, related to the $U(K)$ global
symmetry of the $K$ hypermultiplets. Since the theory on the D5-branes is free, in this limit, only the two-point correlation functions of these modes remain non-zero (if we normalize them to one) and the higher correlators vanish. Note that if we use the standard normalization for these operators, their 2-point functions and other correlators would scale with $N$
and diverge in this limit.

Another option is to take a large $N$ limit while keeping $g_{\rm st} N$ fixed. The difference
is that in this case also the massive open string modes living on the $K$ D5-branes
remain at finite mass, though they are still free (recall that
$R_{\rm AdS}^2 / \ell_{\rm st}^2 \simeq \sqrt{g_{\rm st} N}$). So, the decoupled sector
on the gravitational side contains the full free theory of open strings living on D5-branes on $AdS_4\times \mathbb{S}^2$.
On the CFT side, in such a limit, the relevant sector of the theory on the defect would contain not just the chiral operators mentioned above but also many non-chiral operators that are dual to the open string modes. However, it is still true that when we scale the two-point functions of these operators to one, all higher correlation functions vanish.

\subsubsection{NS5-branes}

Consider next taking NS5-branes instead of D5-branes, wrapping the same space. The interpretation in the four-dimensional field theory is now rather different (though it is by construction related to the previous case by S-duality). For a single NS5-brane, one has a $2+1$
dimensional defect such that on each side of the defect we have the four-dimensional ${\cal N}=4$
$SU(N)$ super Yang-Mills theory, and on the defect we have a bifundamental hypermultiplet
linking the two sides. For $K$ NS5-branes, the theory living on the defect is more complicated. It can be 
defined as the IR limit of the quiver theory that is obtained by separating the NS5-branes.

Again, at low energies on the NS5-branes, we have a six-dimensional $U(K)$ super Yang-Mills theory on $AdS_4\times \mathbb{S}^2$, but now its gauge coupling is given by
\be
{R_{\rm AdS}^2 \over g_{\rm YM}^2} \simeq {R_{\rm AdS}^2 \over \ell_{\rm st}^2} \simeq \sqrt{g_{\rm st} N}.
\ee

For $N\to \infty$ with $g_{\rm st}$ fixed, we again obtain a free six-dimensional $U(K)$ super Yang-Mills theory as before. However, if we now take $N\to \infty$ with fixed $g_{\rm st} N$, the $U(K)$ gauge coupling remains finite. In fact, since the string scale is kept fixed, this decoupling limit gives a non-trivial ``little string theory'' \cite{Berkooz:1997cq,Seiberg:1997zk,Maldacena:1997cg,Aharony:1998ub} with
${\cal N}=(1,1)$ supersymmetry, living on $AdS_4\times \mathbb{S}^2$. This is our first example of a
non-trivial decoupling limit in which higher-point functions are non-zero, and it is also an example of a decoupling limit that does not give a local field theory. We will discuss in more detail a similar example in section \ref{lst} below.

\subsubsection{D1-branes}

As our final example of `rigid holography', consider $K$ D1-branes wrapping an $AdS_2$ subspace of $AdS_5$.
This configuration describes in the field theory a straight 't Hooft line in the $K$'th product
of fundamental representations of (the magnetic dual of) $SU(N)$.

At low energies, we now get a two-dimensional $U(K)$ super Yang-Mills theory on $AdS_2$, with gauge
coupling in AdS units
\be
{R_{\rm AdS}^2 g_{\rm YM}^2} = {{R_{\rm AdS}^2 g_{\rm st}} \over  {\ell_{\rm st}^2}} \simeq \sqrt{N g_{\rm st}^3}.
\ee
If we now take $N\to \infty$ with fixed $g_{\rm st}$ the two-dimensional gauge theory goes to infinite coupling. But now we have another interesting possibility of taking $N\to \infty$ keeping
$N g_{\rm st}^3$ fixed. In this limit, the massive string modes still become infinitely massive and decouple, and we are left with the $(1+1)$ dimensional super Yang-Mills theory on $AdS_2$
with a finite gauge coupling. So this gives us a first example where we have a decoupled
non-trivial local field theory on $AdS$ space, arising as a decoupled sector in a CFT.

\subsection{The Decoupling Limit of $AdS_5 \times \mathbb{S}^5/\mathbb{Z}_K$}

The situation we described in the previous sections, where we study the six-dimensional ${\cal N}=(2,0)$ $A_{K-1}$ theory living on $AdS_5\times \mathbb{S}^1$, is an example of a defect in the bulk. Instead of having branes, we now have an orbifold singularity, and modes localized at the singularity.
The discussion of this case is parallel to the discussion of branes. Indeed, there are
various cases where string dualities interchange modes localized at defects with modes living on branes.


In general, the six-dimensional ${\cal N}=(2,0)$ theory that we have at the singularity is not decoupled from the other modes in the bulk (at the orbifold point, the twisted sector is not decoupled from the untwisted sector). However, as in our previous examples, we can decouple it if we take a large $N$ limit. In our case, we can take large $N$ with fixed $g_{\rm st}$ such that the bulk physics becomes free, and such that we decouple the massive string modes on the orbifold singularity. In addition, as discussed above, we need to take $(K-1)$ of the gauge couplings of $SU(N)^K$ to infinity. In order to remain at a specific point in the `moduli space' of the six-dimensional ${\cal N}=(2,0)$ theory, we need to scale the integrals of the $B_2$ field over the vanishing 2-cycles so that the tension of the wrapped D3-branes remains finite (measured in units of the AdS radius). In this limit,  the six-dimensional ${\cal N} = (2,0)$ theory on $AdS_5\times \mathbb{S}^1$ is `dual' to a sector of the four-dimensional ${\cal N}=2$ SCFT with $\widehat{A}_{K-1}$ quiver gauge group. As we discussed above, in the S-dual picture, we know that this sector contains the $SU(K)\times P(K)$ theory and also some part of the $Q(N,K)$ theory. We expect that this sector should include all operators that remain with nontrivial correlation functions in the large $N$ limit, though we could not find a way to characterize this sector more precisely. For large $K$, we know from the properties of the ${\cal N}=(2,0)$ theories that these correlation functions should be suppressed by powers of $K$, but it is not clear how to see this directly in the four-dimensional SCFT.
Note that, if we sit at the origin of the `moduli space', the four-dimensional $SU(K)$ gauge theory is free (and there it does contain a conserved energy-momentum tensor of its own). If we sit at other points in the `moduli space', the different four-dimensional theories couple to each other and it is not obvious that there is any local conserved energy-momentum tensor associated with the decoupled sector.
\section{Universality}
\label{universality}

In the previous sections, we discussed how we can study the six-dimensional ${\cal N}~=~(2,0)$ $A_{K-1}$ theory on $AdS_5\times \mathbb{S}^1$ by embedding it in type IIB string theory on $AdS_5\times {\mathbb S}^5/\mathbb{Z}_K$, and how we can describe it as a decoupled sector in the large $N$ limit of the holographic dual four-dimensional ${\cal N}=2$ SCFT with $SU(N)^K$ quiver gauge group. This embedding is by no means unique or special. There is a variety of other equally viable string backgrounds with the same amount of supersymmetry that also involve the six-dimensional ${\cal N}=(2,0)$ theory on $AdS_5\times \mathbb{S}^1$ in certain low-energy limits. Here, we will analyze a specific class of these backgrounds and learn that they all lead to similar physics to the one described in the previous sections. In particular, they involve the same boundary conditions for the gauge fields on $AdS_5\times \mathbb{S}^1$. This implies that the same six-dimensional theory arises universally as a decoupled sector in a variety of different four-dimensional large-$N$ gauge theories, in a limit similar to the decoupling limit that we described in the previous section. 

A class of additional models whose gravitational dual includes the six-dimensional ${\cal N}=(2,0)$ $A_{K-1}$ theory on $AdS_5\times \mathbb{S}^1$ arises from brane constructions in Type IIA string theory, with $N$ D4-branes intersecting and ending on $K$ NS5-branes and any number of D6-branes. At low-energy, they give rise to a four-dimensional ${\cal N}=2$ SCFT. The theory discussed in the previous sections can be viewed as a `circular quiver' of this type with no D6-branes, which arises when the background is compactified on a circle along a direction on the D4-branes perpendicular to the NS5-branes. 
We can generalize this to many SCFTs that arise from `linear quivers' of this type, with D6-branes giving hypermultiplets in the fundamental representation that are needed to ensure quantum conformal invariance. Focusing on cases which have a type IIA description, the gravitational duals of these theories were constructed in \cite{Aharony:2012tz} (following \cite{Gaiotto:2009gz,ReidEdwards:2010qs}), and they involve type IIA string theory on a warped product of $AdS_5$ times $\mathbb{S}^1\times \mathbb{S}^2$ times a Riemann surface ${\cal M}_2$. The background arising from a brane configuration with $K$ NS5-branes includes $K$ NS5-branes wrapping $AdS_5\times \mathbb{S}^1$, and these branes come together when we take the coupling constants of the four-dimensional ${\cal N}=2$ gauge theories to infinity. Thus, in this limit, which is similar to the strong coupling limit we took in the previous sections, the gravitational dual contains at low energies the six-dimensional ${\cal N}=(2,0)$ $A_{K-1}$ theory on $AdS_5\times \mathbb{S}^1$. Note that also in this Type IIA embedding the radii of $AdS_5$ and of $\mathbb{S}^1$ are equal (though here they
vary in the geometry).

We can again  study the strong coupling limit of the four-dimensional ${\cal N}=2$ SCFT using its Gaiotto description in terms of punctured Riemann surfaces. 
This time, the punctured Riemann surface has the topology of a sphere, and it has additional types of punctures that are related to the D6-branes. Nevertheless, for theories arising from a brane configuration with $K$ NS5-branes, there are always $K$ minimal punctures associated with $U(1)$ global symmetries, and these punctures come together in the strong coupling limit. The description of this strong coupling limit is thus almost identical to what we had above: again the coupling constant of the S-dual $SU(K)$ theory goes to zero when we go to the origin of the appropriate `moduli space', but this time it is coupled to $P(K)$ times a different SCFT with an $SU(K)$ global symmetry. So, we can repeat the whole analysis above for all of these cases, namely, for the theories defined by the Riemann surface with $K$ minimal punctures and any number of extra punctures of {\sl any} type. In particular, we can have any number of D6-branes and any number of D4-branes ending on each D6-brane, as long as we maintain the quantum conformal invariance by ensuring vanishing of all the beta functions \footnote{The dual supergravity backgrounds contain different configurations of D6-branes wrapping $AdS_5\times \mathbb{S}^2$, but these are separated in the compact space from the stack of $K$ NS5-branes  \cite{Aharony:2012tz}.}. The facts that the background in this case is a warped product instead of a direct product as above, and that NS5-branes in type IIA are involved instead of a singular orbifold in type IIB, do not make any real difference. The arguments above suggest that all of these four-dimensional ${\cal N}=2$ theories contain a decoupled sector in the large $N$ limit which is identical, and is the holographic `dual' of the six-dimensional ${\cal N}=(2,0)$ $A_{K-1}$ theory on $AdS_5\times \mathbb{S}^1$.

As an illustrative example, consider the brane construction involving D4-branes intersecting and ending on $K$ NS5-branes (for even $K$), with $2N$ additional D6-branes intersecting the D4-branes in the middle, which corresponds to the
\be
SU(N)
\times \cdots \times SU((K/2-1)N) \times [SU(KN/2)]_{\oplus_{N_f} H} \times SU((K/2-1)N) \times \cdots \times SU(N)
\nonumber
\ee
linear quiver gauge theory, with $N_f = 2N$ extra hypermultiplets in the fundamental representation coupled to the middle $SU(KN/2)$ gauge group. The gravitational dual in this case contains $K$ NS5-branes, and it is weakly curved when $K$ and $N$ are both large and the coupling constants are taken to infinity \footnote{There are also weakly curved backgrounds if all the couplings except a small number are taken to infinity, and these contain several stacks of separated NS5-branes \cite{Aharony:2012tz}.}. Gaiotto's description of these linear quiver gauge theories involves a sphere with $K$ minimal punctures and two maximal punctures (corresponding to $SU(N)$ global symmetries). Our arguments imply that these theories also contain a decoupled sector `dual' to the ${\cal N}=(2,0)$ $A_{K-1}$ theory on $AdS_5\times \mathbb{S}^1$, just like the circular quiver gauge theories discussed in the previous sections. A particularly simple case arises for $K=2$, where we just have the $SU(N)$ gauge theory with $N_f=2N$ hypermultiplets in the fundamental representation. In this case,  the dual gravitational background is highly curved (since it includes the near-horizon region of just two NS5-branes), but we do not expect this to affect the low-energy theory living on the NS5-branes, so we claim that this ${\cal N}=2$ $SU(N)$ gauge theory with $N_f=2N$ still includes a decoupled sector in the large $N$ limit corresponding to the ${\cal N}=(2,0)$ $A_1$ theory on $AdS_5\times \mathbb{S}^1$. The strong coupling limit of this class of theories indeed develops a weakly coupled $SU(2)$ gauge group as in our discussion above  (see, for example, section 12.3.1 of the review \cite{Tachikawa:2013kta} and the original references therein).

A similar picture arises for more general theories related to six-dimensional SCFTs on Riemann surfaces with $K$
minimal punctures. General theories of this type have M theory gravitational duals constructed in \cite{Maldacena:2000mw,Gaiotto:2009gz}, and it seems
that in the limit where these $K$ punctures come together, we have $K$ M5-branes wrapping $AdS_5\times \mathbb{S}^1$ coming together, leading to a low-energy ${\cal N}=(2,0)$ $A_{K-1}$ theory on $AdS_5\times \mathbb{S}^1$. It is thus natural to
put forward the following universality conjecture: 
\vskip0.5cm
\noindent
{\it Any four-dimensional ${\cal N}=2$ SCFT of class ${\cal S}$ whose Gaiotto description contains
a Riemann surface with $K$ minimal punctures (and any number of additional punctures), and that has
a strong coupling limit which brings together the $K$ regular punctures, contains (in this limit) a decoupled sector 
in the large $N$ limit that is dual by `rigid holography' to the six-dimensional ${\cal N}=(2,0)$ $A_{K-1}$ theory on $AdS_5\times \mathbb{S}^1$.}
\vskip0.5cm

Our discussion above implies that this decoupled sector always has a description in which it contains the four-dimensional $SU(K)\times P(K)$ theory. This theory is presumably coupled to a universal subsector in the remainder of the four-dimensional SCFT. Our conjecture implies that also all the various SCFTs that have an $SU(K)$-type puncture, that arise (coupled to the gauge group $SU(K)$) in these constructions, have a universal sector that decouples in the large $N$ limit.
We expect that, as above, in the simplest description, the $SU(K)\times P(K)$ theory lives on the boundary of $AdS_5$.

\section{Microstates and Thermodynamics}
\label{section7}

One mysterious aspect of the six-dimensional ${\cal N}=(2,0)$ theory is that its number of elementary degrees of freedom, as probed by the six-dimensional conformal and R-symmetry anomalies and by the thermal free energy computed from the AdS/CFT correspondence, scales very differently from that of quantum field theories that have weakly coupled limits. For instance, consider the free energy of the $A_{K-1}$ theory when the theory is put on a finite space ${\cal M}$ (this can be for instance $\mathbb{T}^5$) and coupled in the canonical ensemble to a heat bath of temperature $T$. Using the AdS/CFT correspondence, it was found that for large $K$ the free energy at temperature $T$ is proportional to a coefficient scaling as $O(K^3)$ times $T^6 \mbox{Vol}({\cal M})$. Note that the dependence on the temperature and on the volume is a consequence of the conformal invariance of the theory.

In the previous sections, we saw that the boundary conditions for our six-dimensional theory on $AdS_5\times \mathbb{S}^1$ that arise naturally from string theory couple it to a four-dimensional theory on the boundary that also has $O(K^3)$ degrees of freedom (as measured by its four-dimensional conformal and R-symmetry anomalies). It is natural to inquire whether there is any relation between these two apperances of $K^3$, and whether this can shed any light of the counting of degrees of freedom for the six-dimensional theory. In this section, we will try to answer this question by analyzing the
conformal anomaly of the four-dimensional, S-dual ${\cal N}=2$ SCFT and the thermodynamics of the $A_{K-1}$ theory on $AdS_5\times \mathbb{S}^1$. We will see that at least in this context there does not seem to be any relation between the two appearances of $K^3$.

\subsection{Conformal Anomaly}
We first compute the conformal anomalies of the various sectors that our four-dimensional SCFT decomposes to in the singular limit, since this gives a way to count the number of degrees of freedom. In a free four-dimensional ${\cal N}=2$ theory, the anomalies
$a$ and $c$ are linear combinations of the number of hypermultiplets $n_H$ and the number of vector multiplets $n_V$, so
we can characterize them by these numbers. The original $SU(N)^K$ quiver theory that we started from has
\begin{equation} \label{anomone}
SU(N)^K : \qquad n_V = K(N^2-1), \qquad \qquad n_H = K N^2.
\end{equation}
The anomalies of the $P(K)$ theory are independent of its parameters, so we can compute them using its limit as a free gauge theory (\ref{new_gauge}) :
\be
P(K) : \qquad && n_V = \sum_{n=2}^{K-1} (n^2-1) = \frac{1}{6} (K-1) (K-2) (2K+3), \nonumber \\ 
&& n_H = \sum_{n=2}^K n(n-1) = \frac{1}{3} K (K+1) (K-1). \label{anomtwo}
\ee
Note that in the large $K$ limit the conformal anomalies of the $P(K)$ theory scale as $K^3$; this is reminiscent of the number of degrees of freedom of the six-dimensional $A_{K-1}$ theories, and we will return to this similarity below.
The fact that the anomalies are independent of the parameters means that those of the $Q(N,K)$ theory are given
by subtracting from (\ref{anomone}) the sum of (\ref{anomtwo}) and a free $SU(K)$ gauge theory, giving effective values
\be \label{anomthree}
Q(N,K) : \qquad && n_V = K(N^2-1) - \frac{K^3}{3} - \frac{K^2}{2} + \frac{5K}{6} - 1, \nonumber \\
&& n_H = K N^2 - \frac{1}{3}(K^3-K).
\ee
This can also be computed directly using the rules for computing anomalies of class ${\cal S}$ theories, giving the same result. Still, we do not know how many of the degrees of freedom  measured in \eqref{anomthree} remain as part of the decoupled theory in the large $N$ limit but our proposal asserts that they ought to be independent of $N$. 

\subsection{Thermodynamics}
Let us next turn to the bulk and study thermodynamics. In order to avoid infrared divergences, in this section, we will in global coordinates for $AdS_5$ so that its boundary is $\mathbb{S}^3\times \mathbb{R}$ or, at finite temperature, $\mathbb{S}^3\times \mathbb{S}^1$ with the $\mathbb{S}^1$ of circumference $\beta = 1/T$. The $\mathbb{S}^3$ has a radius $L$, and conformal invariance implies that the thermodynamical behavior of the theory depends on the dimensionless combination $T L$. 

As we discussed above, 
the states that correspond to the six-dimensional ${\cal N}=(2,0)$ $A_{K-1}$ theory on $AdS_5\times \mathbb{S}^1$ arise as a subsector of the states of the four-dimensional ${\cal N}=2$ $SU(N)^K$ SCFT at the singular point in its parameter space. So we will start by studying the thermodynamics of the latter theory.
For simplicity, we will take $K$ to be large but still parametrically smaller than any power of $N$:
\be
N^\alpha \gg K \gg 1 \qquad \mbox{where} \qquad \alpha >0.
\label{limit}
\ee

In the canonical ensemble, large $N$ gauge theories on $\mathbb{S}^3$ undergo phase transitions at a temperature of order $T \sim 1/L$
\cite{Witten:1998zw,Aharony:2003sx}. Below the phase transition, the free energy is of order one. Above it, the free energy is of the order of the number of fields. In particular we expect this to be true for the $SU(N)^K$ quiver gauge theory, which at high temperatures should have a free energy of order $K N^2$ at any point in its parameter space.
Near the singular point, we saw that this theory decomposes into a weakly interating $SU(K)$ gauge theory coupled to $Q(N,K)\times P(K)$. For large $K$, we expect that there will still be a similar transition (which would perhaps split into separate transitions for the different factors).
For $T \ll 1/L$, the free energy of all these theories is $O(1)$ (and can have a complicated dependence on the temperature times the radius). For $T \gg 1/L$, the free energy is proportional to the volume of $\mathbb{S}^3$ and has contributions of order $K^3$ for the $P(K)$ theory and of (much larger) order $K N^2$ for the $Q(N,K)$ theory.
Most of the states of the $Q(N,K)$ theory are dual to bulk states that have nothing to do with the six-dimensional ${\cal N}=(2,0)$ theory. It is not clear how many of the states in the $Q(N,K)$ theory remain in the decoupled sector which describes purely the six-dimensional ${\cal N}=(2,0)$ theory. Nevertheless, we expect that this number will not scale with $N$. Alternatively, we can isolate the relevant part of the $Q(N,K)$ theory by considering the derivative of its free energy with respect to $K$. It should be stressed that the decoupled sector, even though it is generally not a local CFT, still has a state-operator correspondence.

For our purposes, we are interested in the regime of $T \gg 1/L$, where we can see the $K^3$ degrees of freedom of the $P(K)$
theory. However, unfortunately, this regime is not related to the thermodynamics of the ${\cal N}=(2,0)$ theory on $AdS_5\times \mathbb{S}^1$. In string theory on $AdS_5\times \mathbb{S}^5/\mathbb{Z}_K$, the phase transition described above, at $T \sim 1/L$, maps into the 
Hawking-Page transition. For $T \ll 1/L$ the bulk physics is dominated by a thermal AdS space, while for $T \gg 1/L$ it is dominated by an AdS black hole \cite{Hawking:1982dh,Witten:1998zw}. This means
that, at least as long as we do not take the strict decoupling limit, the dominant background for $T \gg 1/L$ is not a thermal
AdS space, so we cannot directly relate the $T \gg 1/L$ behavior of the four-dimensional CFT to thermodynamics of a theory on AdS space.

We can get around this problem by looking at the microcanonical ensemble since there, near the singular
point, we can separate the contributions of the $P(K)$ theory from the contributions of generic states in the $Q(N,K)$ theory.
At energies far below $K^2/L$,  the entropy of both theories is of order one. At energies much higher than
$K^2/L$ (but much smaller than $N^2/L$), the $P(K)$ theory has a contribution to the entropy density that is proportional to its number of degrees of freedom, $O(K^3)$ (this is clear when this
theory is weakly coupled, but we expect it to be true for generic points in its parameter space).
At these energies, this is the leading contribution to the microcanonical ensemble in our limit (\ref{limit}), while at higher energies (above $N^2/L$) it is swamped by generic states of the four-dimensional large $N$ SCFT.
%


Next, let us try to understand the thermodynamics from the point of view of string theory on $AdS_5\times \mathbb{S}^5/\mathbb{Z}_K$, where there are several different contributions to the thermodynamics.
The bulk fields have a similar behavior to the bulk fields
in the $AdS_5\times \mathbb{S}^5$ dual of the four-dimensional ${\cal N}=4$ super Yang-Mills theory, up to some factors of $K$. The fields living at the singularity at low energies, 
of order $O(1/R_{\rm AdS})$ in the bulk (translated to $O(1/L)$ from the point of view of the dual four-dimensional SCFT), give a gas of light particles (including the strongly coupled
five-dimensional $SU(K)$ gauge fields). At energies much higher than $1/R_{\rm AdS}$, and certainly above $K^3/R_{\rm AdS}$, 
thermal excitations on $AdS_5\times \mathbb{S}^1$ are insensitive to the boundary conditions, and the bulk thermodynamics should contain that of the six-dimensional ${\cal N}=(2,0)$ SCFT on the space $AdS_5\times \mathbb{S}^1$. Its free energy behaves like that of a six-dimensional theory with $K^3$ degrees of freedom, with an effective volume of order $R_{\rm AdS}^4 R_{S^1}$, since anti-de Sitter space behaves as a box.

We already discussed above the canonical ensemble. At low temperatures, we have a gas of particles both inside the bulk and on the boundary which, by construction, reproduces the low-energy phase of the four-dimensional field theory. The states corresponding to the $SU(K)\times P(K)$ theory live on the boundary of $AdS_5$, while the excitations of the bulk fields living at the singularity are dual to excitations (generally charged under $SU(K)$, since this charge can be canceled by boundary fields) involving the strongly interacting $Q(N,K)$ theory. In the full string theory, above a temperature of order $1/L$,  we have a Hawking-Page transition to an AdS-black hole in the bulk. So, we no longer have our six-dimensional theory in the bulk living on $AdS_5\times \mathbb{S}^1$ \footnote{Analyzing this theory on the AdS black hole background is interesting but beyond the scope of this paper.}.

However, in the microcanonical ensemble, we can look for states in the intermediate regime
$K^2/L \ll E \ll N^2/L$ discussed above. In the four-dimensional SCFT, we know that these states give rise to
an entropy going as $K^3 (E L)^{3/4}$, as appropriate for a four-dimensional conformal theory with a number of degrees of freedom of $O(K^3)$. 
Here, it does not matter whether we impose the four-dimensional $SU(K)$ singlet constraint or not.
In our bulk description, we find the same states living on the boundary, but in addition to this we have the
$K^3$ degrees of freedom of the ${\cal N}=(2,0)$ theory in  the bulk. It is tempting to try to identify the two
contributions. However, the bulk contribution gives an entropy going as $K^3 (E L)^{5/6}$, which
is much bigger than the one we find from the boundary $SU(K)\times P(K)$ theories. Presumably, these six-dimensional states are related to the $Q(N,K)$ sector of the CFT, whose density of states we do not know how to compute in this regime. 
So, even if some of the $SU(K)\times P(K)$ boundary states are related to the ${\cal N}=(2,0)$ theory
in the bulk, they do not fully reveal and explain its entropy. In this sense, it is yet unclear from our description how to approach counting the degrees of freedom of the ${\cal N}=(2,0)$ theory.


\section{Little String Theories on $AdS_5\times \mathbb{S}^1$}
\label{lst}

Up to now, we analyzed Type IIB string theory on $AdS_5\times \mathbb{S}^5/\mathbb{Z}_K$, which contains at low energies (in the singular limit of the orbifold) the six-dimensional ${\cal N}=(2,0)$ theory on $AdS_5\times \mathbb{S}^1$. We saw that we could use this to learn about the ${\cal N}=(2,0)$ superconformal theory and to identify a decoupled sector in the large-$N$ limit of the corresponding four-dimensional ${\cal N}=2$ quiver SCFTs. In order to isolate the properties of the six-dimensional ${\cal N}=(2,0)$ theory, we took the string scale and the Planck scale to infinity while keeping fixed the AdS curvature scale and, when we are slightly away from the singular point, the tensions (in units of the AdS radius) of the BPS strings arising from wrapped D3-branes. It is important to note that our discussions did not rely on the fact that the decoupled sector, the six-dimensional ${\cal N}=(2,0)$ theory, itself was conformal. So, we can also try to extend our discussions to the regime that the decoupled theory on $AdS$ is non-conformal, while still being decoupled from the bulk gravity and realizing the rigid holography. 



\subsection{Isolating Little String Theory}
There is an interesting class of theories that are known to be decoupled from gravity, which also arise as a decoupling limit of Type IIB string theory on the singular orbifold $\mathbb{C}^2/\mathbb{Z}_K$. Take a limit of the Type IIB string theory in which, by sending the string coupling $g_{\rm s}$ to zero, the string scale remains fixed while the Planck scale goes to infinity. As reviewed in \cite{Aharony:1999ks}, this limit decouples gravity and leads to an interesting {\sl nonlocal} quantum theory called `little string theory' (LST). This theory still preserves ${\cal N}=(2,0)$ supersymmetry in six dimensions. At low energies below the string scale, this theory reduces to the ${\cal N}=(2,0)$ SCFT, but its high-energy behavior is more like that of a string theory. In particular, the theory has, even at the origin of its moduli space, a BPS string with finite tension arising from the original Type IIB fundamental string. On $\mathbb{R}^{5,1}$, these quantum theories have interesting properties such as T-duality on circles and a continuous spectrum above a mass gap. 

Despite prolonged efforts, many of the novel properties of the LST are still not well understood. As such, any new ideas and approaches would be most welcome. 
Here, we propose such an approach. A variation of our discussion in the previous sections allows us to take a different decoupling limit of Type IIB string theory on $AdS_5\times \mathbb{S}^5/\mathbb{Z}_K$ that will lead to the ${\cal N}=(2,0)$ LST on $AdS_5\times \mathbb{S}^1$, instead of just the low-energy ${\cal N}=(2,0)$ SCFT. In Type IIB string theory, the ratio of the string scale to the AdS scale is given by $R_{\rm AdS}^4 / \ell_{\rm s}^4 = g_{\rm s} N = \lambda$ (ignoring some multiplicative factors of $K$, that is always kept fixed). Previously, we took this to infinity in the large $N$ limit. We can alternatively keep this ratio fixed. To do so, we need to scale the string coupling $g_{\rm s}$ to zero as $1/N$ in the large $N$ limit. Note that the LST on $AdS_5\times \mathbb{S}^1$ has an extra dimensionless parameter compared to the low-energy limit ${\cal N}=(2,0)$ SCFT on $AdS_5\times \mathbb{S}^1$, which is the tension of its extra BPS string (in units of the AdS radius $R_{\rm AdS}$)
\be
T_{\rm F} R_{\rm AdS}^2 = {R_{\rm AdS}^2 \over 2 \pi \ell_s^2} = {\sqrt{\lambda} \over 2 \pi}.
\label{fstringtension}
\ee
We also want to keep the tensions of the BPS strings arising from wrapped D3-branes
\be
T^{(a)}_{\rm D3} R_{\rm AdS}^2 = {B^{(a)}_2 \over 2 \pi g_{\rm s} \ell_s^2}  R_{\rm AdS}^2 = {B_2^{(a)} N \over 2 \pi \sqrt{\lambda}} \qquad \qquad (a = 1, \cdots, K)
\label{dstringtension}
\ee
finite and of the same order as the tension of the extra BPS string. This is achieved by taking
\be
{1 \over N} \sim g_{\rm s} \sim B_2^{(a)} \mbox{\ with fixed ratios} \qquad \mbox{as} \qquad N \rightarrow \infty.
\label{scalingLST}
\ee
The resulting LST on $AdS_5\times \mathbb{S}^1$ reduces at low energies to the ${\cal N}=(2,0)$ SCFT. In particular, it has the same `moduli space' with the same BPS strings on this `moduli space', but it has an extra BPS string whose tension is independent of the position on the `moduli space'. Note that if one computes the index (the supersymmetric partition function on $\mathbb{S}^3\times \mathbb{S}^1$) of the decoupled sector that we obtain in the four-dimensional ${\cal N}=2$ SCFT, one gets the same answer in this new limit as in our previous sections (the index is independent of $g_s$). So the LST does not contain any extra BPS particles that are measured by this index, but it does contain an extra BPS string.

We can further take the double-scaled little string theory (DSLST) limit \cite{Giveon:1999zm,Giveon:1999px,Giveon:1999tq}, such that its holographic dual is weakly coupled. This is achieved by taking
$T_{D3}^{(a)} \gg T_F$:
\be
{1 \over N} \sim g_{\rm s} \ll B_2^{(a)} 
\qquad
\mbox{as} \qquad N \rightarrow \infty,
\label{scalingDSLST}
\ee
with a finite large ratio $B_2^{(a)} / g_s$.

The (DS)LST we obtain in the decoupling limit has ${\cal N}=(2,0)$  supersymmetry and is defined over the spacetime $AdS_5 \times \mathbb{S}^1$. Utilizing the T-duality symmetry of LST, we can also  obtain the (DS)LST with ${\cal N} = (1,1)$ supersymmetry, now defined over the spacetime $AdS_5 \times \widetilde{\mathbb{S}}^1$. Here, the radius of $\widetilde{\mathbb{S}}^1$ is $1/R$ in the unit of the LST string scale.

\subsection{Rigid Holography}
The different scaling \eqref{scalingLST} necessary to isolate the LST should be reflected in the Gaiotto description of the four-dimensional ${\cal N}=2$ SCFTs. We can identify it as a different scaling for the complex structures. Previously, we just took the $K$ punctures to approach each other on the torus, while keeping the complex structure $\tau_0$ of this torus fixed. Now, we are making the torus very elongated by taking the complex structure $\tau_0 \to i\infty$, in a way that is correlated with $N$, ${\rm Im}(\tau_0) \propto N$. 

\begin{figure}[tbp]
\centering
\vspace*{-2cm}	
\includegraphics[angle=0,width=14cm]{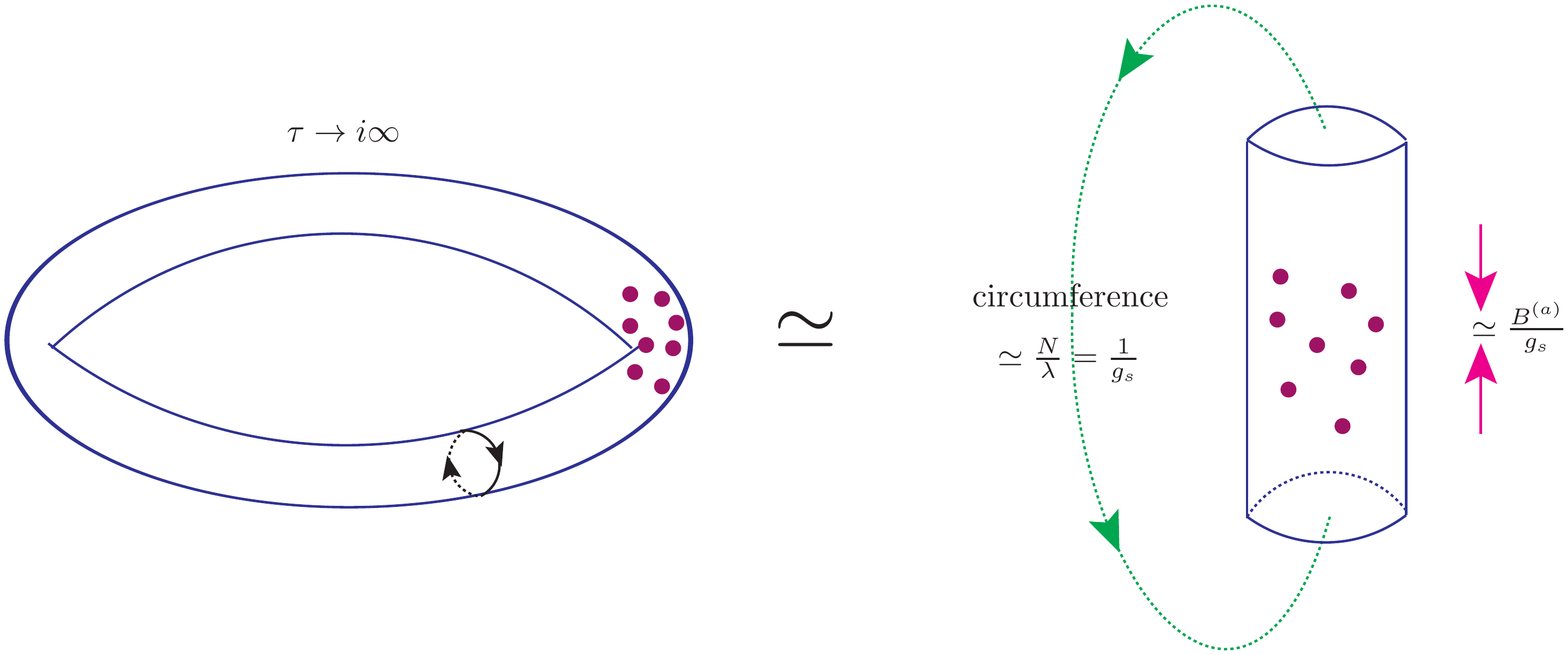}
\vspace*{-1cm}
\caption{\sl The UV curve of the four-dimensional ${\cal N}=2$ superconformal quiver gauge theory that leads in the holographic dual to the little string theory on $AdS_5 \times \mathbb{S}^1$.}
\label{fig:LST-GaiottoCurve}
\end{figure}
%

To ascertain that this is the correct scaling, we should be able to identify a new BPS string state as the LST string. In our previous discussion, the $(K-1)$ BPS strings coming from wrapped D3-branes were related to $(K-1)$ non-trivial homology 1-cycles on the torus that go around the $K$ approaching punctures. The new BPS string can now be similarly related to the `small' cycle of the torus, whose size in our limit is kept of the same order as the distance between the $K$ approaching punctures, see figure \ref{fig:LST-GaiottoCurve}. Recall that the tension of the wrapped D3-branes is given by (\ref{dstringtension}); if we fix the long cycle of the torus to have length one, then the short cycle has length $g_{\rm s}$, the distances between the punctures are of order $B_2^{(a)}$, and we are taking $B_2^{(a)} \sim g_{\rm s}$. From (\ref{fstringtension}) and (\ref{dstringtension}), we see that this scaling puts
\be
T^{(a)}_{\rm D3} \sim T_F \quad \mbox{for all} \quad a= 1, \cdots, K-1.
\ee

Stated differently, in this description, our decoupling limit corresponds to taking $K$ punctures together and also shrinking one of the cycles of the torus, such that locally we have $K$ points on a cylinder (with fixed distances compared to the width of the cylinder, if we are staying at a fixed point in the `moduli space' of the LST with fixed ratios between all the BPS strings). The DSLST limit is obtained by further tuning these ratios hierarchically. 

\subsection{Microstates and Thermodynamics}
Naively, the LST rigid holography appears to contradict microstate countings. The density of states of LST at high energies is believed to grow exponentially, as $\rho(E) \simeq \exp(E \ell_{\rm s})$ \cite{Maldacena:1997cg}, while the high-energy density of states in our four-dimensional ${\cal N}=2$ quiver SCFT grows only as a power of the energy. However, the field theory density has a prefactor of $N^2$ and we decouple the LST only when $N \to \infty$. So, there is no contradiction. 

For any finite $N$, at high enough energies the LST no longer decouples from the bulk physics, and the exponential density of states is cut off. The LST defined on $\mathbb{R}^{5,1}$ has a continuous mass spectrum above a mass gap of order $1/\ell_{\rm s}$.  Here, for the LST defined on $AdS_5\times \mathbb{S}^1$, it seems this is replaced by a discrete spectrum. It would be interesting to study the remnants of the continuous spectrum when the theory is put on $AdS_5\times \mathbb{S}^1$. This continuous spectrum is most easily seen in the holographic description of the LST, when the NS5-branes are replaced by their near-horizon limit. It is an interesting problem for the future to find such a holographic description for the LST on  $AdS_5\times \mathbb{S}^1$, for instance by taking the near-horizon limit of the ${\mathbb Z}_K$ singularity in the background discussed above. We expect this limit to be related to the decoupling limits (\ref{scalingLST}), (\ref{scalingDSLST}) needed to isolate the (DS)LST.

\subsection{Generalizations?}
It is natural to generalize this LST construction to the Type IIA string theory setups that we discussed in section \ref{universality}. However, Gaiotto's description implies that this generalization should not work (unlike the generalization for the low-energy ${\cal N}=(2,0)$ SCFT), since the Riemann surface in this case does not have an extra cycle like the one we used above. In the corresponding type IIA backgrounds, indeed, $g_{\rm s}$ varies and it is more subtle to take in this case a decoupling limit that would keep precisely the ${\cal N}=(2,0)$ LST. Our discussion here suggests that it should be impossible to do this. It would be interesting to verify this by carefully studying the different limits of the string theory backgrounds of \cite{Aharony:2012tz}.
\section{Future Directions}
\label{section9}

In this paper, we set out a new general framework for studying field theories on AdS space, by embedding them into string theory and utilizing the AdS/CFT correspondence. We termed this idea `rigid holography', since it relates a sector of a non-gravitational CFT to a higher-dimensional non-gravitational quantum field theory in the bulk AdS space. 

We used this `rigid holography' to study the six-dimensional ${\cal N}=(2,0)$ $A_{K-1}$ theory on $AdS_5\times\mathbb{S}^1$. Utilizing Gaiotto's description of the four-dimensional ${\cal N}=2$ quiver SCFTs that are dual to the string theory backgrounds in which we embedded this theory, we showed how to identify a decoupled sector in a specific regime of the parameter space of the large $N$ limit of the four-dimensional SCFTs. In particular, this gives a prescription for computing the correlation functions of this six-dimensional theory, at least for operators inserted at infinity. 

We used the rigid holography to uncover several surprising properties of this six-dimensional ${\cal N} = (2,0)$ theory. We conjectured that this new correspondence between the decoupled sectors is universal, involving a universal subsector that appears in the large $N$ limit of many four-dimensional ${\cal N}=2$ SCFTs. Using the proposed construction, we should be able to understand this enigmatic six-dimensional theory from new angles, and also to deepen our understanding of the large $N$ and strong coupling limits of four-dimensional ${\cal N}=2$ SCFTs with quiver gauge groups.

There are many interesting open questions that are raised by our analysis, some of which we are currently investigating. Below, we summarize some of these future directions. 

\subsection{More Examples of Rigid Holography}
In this paper, we discussed in detail just a single example of using the proposed `rigid holography' to learn about quantum field theories on AdS space. There are many possible generalizations of our considerations to other non-gravitational $d$-dimensional field theories and `little string theories' living on $AdS_p\times {\cal M}_{d-p}$, with some $(d-p)$-dimensional manifold ${\cal M}_{d-p}$ and with $p=2,3,\cdots,d$. In particular, rigid holography encompasses quantum theories which have local energy-momentum tensors (such as the six-dimensional ${\cal N}=(2,0)$ theory) and which do not have local energy-momentum tensors (such as LSTs). We already discussed a few of these generalizations in Section \ref{reverse}. 

Clearly, there are many wider possibilities. A relatively simple generalization is to study six-dimensional $D_K$ and $E_K$ ${\cal N}=(2,0)$ SCFTs on $AdS_5\times \mathbb{S}^1$, by realizing them as low-energy limits of other orbifolds of type IIB string theory on $AdS_5\times \mathbb{S}^5$. There is even another way to realize the six-dimensional ${\cal N}=(2,0)$ $A_{K-1}$ SCFT on $AdS_5\times \mathbb{S}^1$, with a different ratio of the radii, by taking the low-energy limit of $K$ M5-branes on $AdS_5\times \mathbb{S}^1$ in the $AdS_7\times \mathbb{S}^4$ background of M-theory. We think this realization corresponds to different boundary conditions, and it would be interesting to understand the similarities and differences from the Type IIB string theory case we studied here. More generally, it would be interesting to understand what we can learn from studying quantum field theories on anti-de Sitter space with various boundary conditions.

\subsection{Rigid Holography Made Precise}
\label{precise}
In the six-dimensional example, we presented the general relation between the parameters of the four-dimensional S-dual ${\cal N}=2$ SCFT and the position on the `moduli space' \eqref{ads5modulispace}, but we did not present a precise mapping. It should be possible to find the precise mapping by computing physical observables from both sides. 

In the four-dimensional ${\cal N}=2$ SCFTs with quiver gauge groups, we can compute the partition function on $\mathbb{S}^4$, Wilson and `t Hooft loops, using the supersymmetry localization technique \cite{baerey}, including the presence of BPS defect operators. The previous exact computations of Wilson loop operators \cite{Rey:2010ry} indicated hierarchically separated BPS 
string tensions in the singular limit \footnote{The analysis at the orbifold point is straightforward. For a more recent study, see for example \cite{Mitev:2014yba}. The hierarchical behavior of the Wilson and `t Hooft loop operators as exact marginal deformations are turned on away from the orbifold point was first initiated in \cite{Rey:2010ry}. }.

Extensions to `t Hooft loops and some correlation functions are straightforward, and the correlation functions of loop operators should be compared with those of the BPS strings in the six-dimensional ${\cal N}=(2,0)$ theory on $AdS_5 \times \mathbb{S}^1$. Equating the two computations, we should be able to extract the relation between the position in the `moduli space' and the couplings of the $SU(K)$ and $P(K)$ theories \cite{reyxie,ArelVlad}. 

It would be interesting to compute the
partition function, the elliptic genus of the Wilson and `t Hooft loops, and the BPS defect operators directly in the six-dimensional ${\cal N}=(2,0)$ theory on $AdS_5 \times \mathbb{S}^1$ using localization, and to compare the results with the exact computations from the four-dimensional ${\cal N}=2$ quiver SCFT side \cite{baerey}.

\subsection{What can be Learned from Rigid Holography}

We argued that a sector in the large $N$ limit of four-dimensional ${\cal N}=2$ SCFTs describes all correlation functions of the six-dimensional ${\cal N}=(2,0)$ theory on $AdS_5\times \mathbb{S}^1$ in which the sources are put at the boundary. 
In addition, the line operators in this sector correspond to surface operators in the ${\cal N}=(2,0)$ theory \cite{Ganor:1996nf} that end on the boundary of $AdS_5$.
However, as a quantum theory that does not involve gravity, there are many more observables in the ${\cal N}=(2,0)$ theory on $AdS_5 \times \mathbb{S}^1$, such as local correlators among operators put anywhere in the interior of the $AdS_5$. 

Can we use the information of the `dual' four-dimensional CFT to reconstruct all local correlation functions of the six-dimensional theory? Are these uniquely determined given the ones with the sources at the boundary? This question is analogous to a standard question in quantum field theory in flat space, which is whether the S-matrix determines the local correlation functions or not. Presumably, there is always some freedom in topological degrees of freedom (which may affect, for instance, the spectrum of local operators, without affecting the spectrum of particles and the S-matrix), and it would be interesting to understand if this is the only freedom or not.


\subsection{Classification of AdS Boundary Conditions}
In most of this paper, we focused on a specific boundary condition for the ${\cal N}=(2,0)$ $A_{K-1}$ theory on $AdS_5 \times \mathbb{S}^1$, which couples it to a $SU(K)\times P(K)$ theory on the boundary, and which preserves the full superconformal symmetry only for a specific ratio of the radii. In section \ref{otherbc}, we discussed the more standard boundary condition, which gives an $SU(K)$ global symmetry and which is possible for any ratio of the radii. 

It would be interesting to classify the full set of boundary conditions that preserve the full superconformal symmetry. For instance, one may expect that it should be possible to put any four-dimensional ${\cal N}=2$ asymptotically-free $SU(K)$ gauge theory on the boundary, for a specific ratio of the radii at which the bulk theory would exactly cancel the four-dimensional $SU(K)$ beta function. This can include both standard matter fields coupled to the four-dimensional $SU(K)$ gauge group and `non-Lagrangian' theories coupled to it. It may also be possible to have hybrid boundary conditions in which we couple the five-dimensional $SU(K)$ gauge theory to a four-dimensional gauge theory whose gauge group is a subgroup of $SU(K)$. Are there any additional possiblities~? Which possibilities can be embedded in string/M theory so that we can use the `rigid holography' to study them and relate them to sectors in a four-dimensional ${\cal N}=2$ SCFT ? Are there any other boundary conditions that make sense for an arbitrary ratio of radii, such that one can take the flat space ${\mathbb R}^5\times \mathbb{S}^1$ limit~? What can be said about boundary conditions that preserve less (super)symmetry~?

\subsection{Duality over the Moduli Space}
We found a new type of duality relating two different descriptions of the six-dimensional ${\cal N}=(2,0)$ theory on $AdS_5\times \mathbb{S}^1$; one which far out on the `moduli space' has a $U(1)^{K-1}$ gauge symmetry in the bulk and no degrees of freedom on the boundary, and another which has an $SU(K)$ gauge symmetry in the bulk and is coupled to a four-dimensional ${\cal N}=2$ $SU(K)\times P(K)$ theory on the boundary. Can one provide more evidence for this duality~? In particular, the duality exchanges some degrees of freedom on the boundary with degrees of freedom in the bulk \footnote{For instance, in the second description,  the $U(1)^{K-1}$ global symmetry acts only on boundary degrees of freedom.}, so it seems like a relative of the AdS/CFT correspondence, except that the correspondence is used here just for some part of the degrees of freedom. Can we use the AdS/CFT correspondence to understand this duality better~? Alternatively, does this duality shed a new light on the AdS/CFT correspondence in a decoupled and somewhat simpler setting that does not involve gravity at all~?

There is one technical question which we left open, related to this and to the precise mapping of the parameters discussed in section \ref{precise}, which should be resolved. When we are far on the `moduli space' \eqref{ads5modulispace}, namely, when the tension of the BPS strings is much larger than the AdS excitation scale, the description of our theory with a $U(1)^{K-1}$ theory in the bulk is weakly coupled at low energies. Thus, the alternative description with a boundary $SU(K)\times P(K)$ theory should be strongly coupled so as not to give rise to any extra light fields. This also follows from the fact that the Wilson loops of this theory, which are identified with the BPS strings in $AdS_5$, are far from their weak coupling value. However, a naive computation of the gauge coupling of the four-dimensional $SU(K)$ gauge theory suggests that these expectation values should equal their free value in the large $N$ limit. Presumably, the resolution of this issue is that there is an extra $N$-dependence, arising from the coupling of the $SU(K)$ gauge theory to the large-$N$ four-dimensional $Q(N,K)$ SCFT, which leads to the theory being strongly coupled at large $N$ even though formally its coupling constant vanishes in the large $N$ limit. It would be interesting to confirm this by finding the precise mapping of parameters, or to find an alternative explanation for our results.



\subsection{Decoupling in the Presence of Additional Punctures}
We argued by the universality conjecture that the same decoupled sector arises in the large $N$ limit of many different four-dimensional ${\cal N}=2$ quiver SCFTs, whose Gaiotto description involves a Riemann surface with $K$ minimal punctures. It would be interesting to find additional arguments for this (including direct field theory arguments) and to find better ways to characterize the decoupled sector and the class of theories from which it arises. A naive interpretation of the Gaiotto description for this process would imply that perhaps the Riemann surface decompactifies, so that the decoupled sector can be described in this picture as coming from the six-dimensional ${\cal N} = (2,0)$ $A_{N-1}$ theory on a non-compact surface which is asymptotically a plane with $K$ minimal punctures. It would be very interesting to make this more precise.

It is natural to conjecture that such a decoupling limit exists also for other types of punctures coming together; these also lead by Gaiotto dualities to `throats' coming out of the Riemann surface (see \cite{Tachikawa:2013kta,Gaiotto:2014bja} for recent reviews), similar to the one we found. In general, these `throats' would involve also strongly coupled four-dimensional ${\cal N}=2$ quiver SCFTs and may not have a weakly coupled limit as in the example that we discussed. A priori, one may think that, in the decoupling limit that we discussed in section \ref{reverse}, this would lead to new six-dimensional ${\cal N}=(2,0)$ theories. However, this seems unlikely. Since in our case we found that the `throat' just gave a field theory living on the boundary that coupled to the bulk ${\cal N}=(2,0)$ theory, a more natural suggestion is that different choices of punctures give rise to different boundary conditions for the same ADE-type six-dimensional ${\cal N}=(2,0)$ theories. Presumably, these boundary conditions would generally include a coupling to strongly coupled four-dimensional ${\cal N}=2$ theories at the boundary of $AdS_5$. This issue should be studied further; it should be understood precisely which six-dimensional ${\cal N}=(2,0)$ theory and which boundary conditions arise for each collection of punctures. It would be interesting to understand whether this construction provides a full classification of the possible boundary conditions preserving the $AdS_5$ symmetry and 16 supercharges for the six-dimensional ${\cal N}=(2,0)$ superconformal theories, or if there are other possible boundary conditions which do not arise as decoupled sectors in any four-dimensional ${\cal N}=2$ SCFT.

\subsection{Variations of General Little String Theories}
One can also study limits in which $K$ punctures approach each other, while also taking to zero the size of various cycles  of the Riemann surface and bringing them together with the punctures (as in our discussion of `little string theory' in section \ref{lst}). For example, one could take a handle in a higher genus Riemann surface $\Sigma_g$ and take it to approach the punctures (while scaling its size to be of the same order as the distance between the punctures). As in our discussion in section \ref{lst}, this would lead to a theory with additional BPS strings associated with the cycles of the handle that is brought in. 

\begin{figure}[tbp]
\centering
\vspace*{-1cm}	
\includegraphics[angle=0,width=12cm]{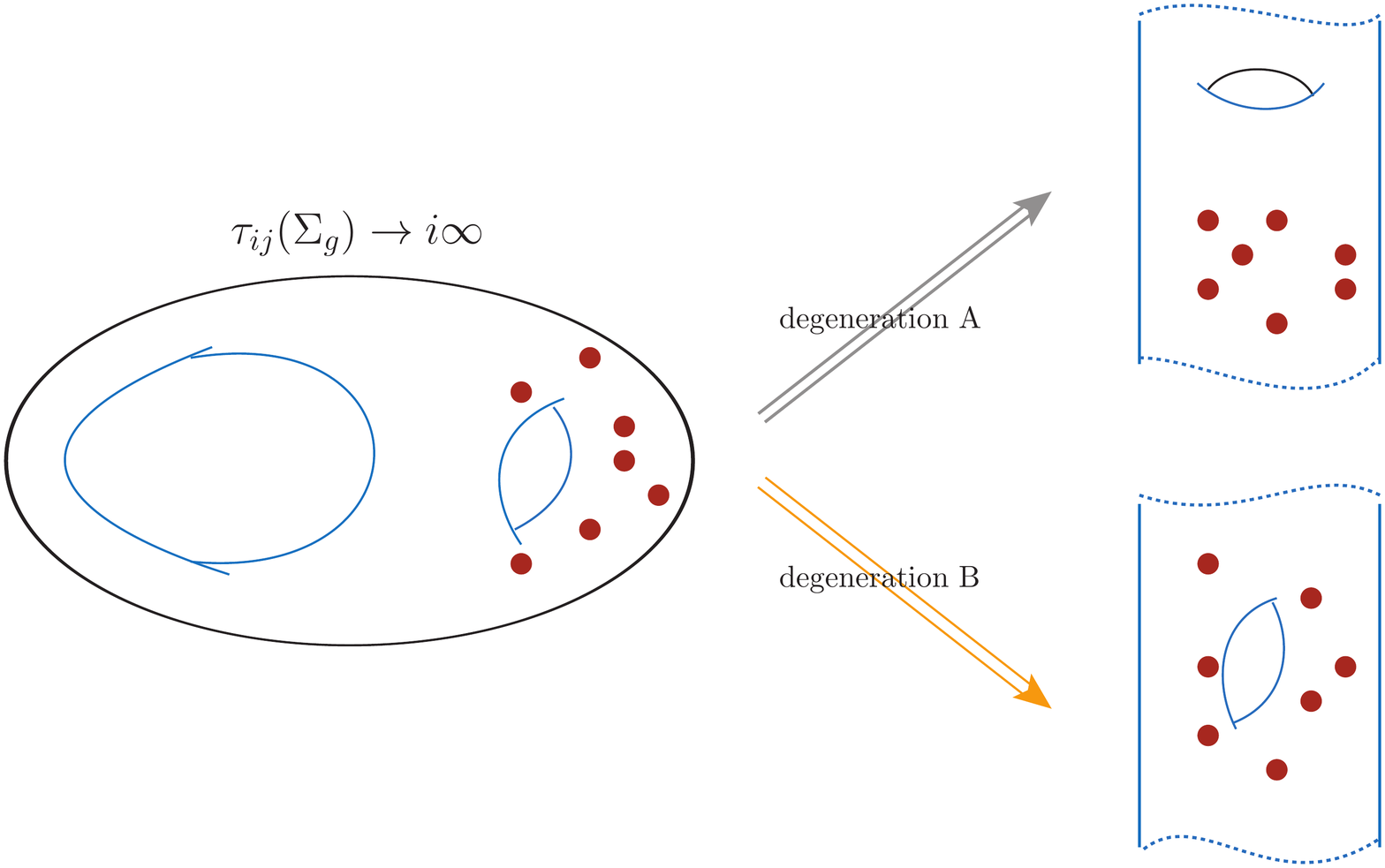}
\vspace*{0cm}
\caption{\sl LST decoupling limits with a handle attached. On the right, cylindrical sections refer to the decoupled `throat', whose two ends are connected by infinitely long tube. The handle may provide an effective puncture (degeneration A) or the handle may be part of the `throat'.}
\label{fig:LST-General-Curve}
\end{figure}
%

Intuitively, it seems that there are different ways to think about such theories with a handle attached, see Figure \ref{fig:LST-General-Curve}. 
If we take the handle to live somewhere in the decoupled `throat' of the Riemann surface, as in degeneration B, then our discussion suggests that it is natural to think about it as modifying the four-dimensional ${\cal N}=2$ quiver SCFTs that live on the boundary (which in this case would be some strongly coupled fixed point). On the other hand, if we keep the handle outside the `throat' but still in the decoupled region, as in degeneration A, it seems more natural to think of it as related to the bulk physics, and perhaps leading to some generalization of `little string theories' which have more than one BPS-saturated string. These two different descriptions are related by Gaiotto-type dualities, with each one being more natural for different ranges of the parameters. This may be similar to the duality between the two different descriptions that we discussed in section \ref{equivalence}. So far, very few LSTs with six-dimensional ${\cal N}=(2,0)$ supersymmetry are known to exist. It would be interesting to understand, following our discussion, if there are actually more waiting to be discovered. 

\subsection{Holography of Rigid Holography}
In this paper, we used the AdS/CFT correspondence only in one direction which we called `rigid holography', by embedding the six-dimensional ${\cal N}=(2,0)$ $A_{K-1}$ theory on $AdS_5\times \mathbb{S}^1$ into a rigid gravitational background, and mapping it to a sector of a four-dimensional ${\cal N}=2$ quiver SCFT. However, in the large $K$ limit, one could also try to invoke holography in the opposite direction, namely, to study the six-dimensional ${\cal N}=(2,0)$ $A_{K-1}$ theory on $AdS_5\times \mathbb{S}^1$ by finding a seven-dimensional gravitational dual for it, generalizing the M theory on $AdS_7\times \mathbb{S}^4$ dual that describes this theory on $\mathbb{R}^{5,1}$ \cite{Maldacena:1997re}. Such a dual would depend on the precise boundary conditions, as in the analogous case of the four-dimensional ${\cal N}=4$ super Yang-Mills theory on $AdS_4$ \cite{Aharony:2010ay,Aharony:2011yc}. As reviewed in Appendix \ref{appendixA}, the product space $AdS_5\times \mathbb{S}^1$ with equal radii is conformally equivalent to the flat space $\mathbb{R}^{5,1}$, so the solution should locally be equivalent to M theory on $AdS_7\times \mathbb{S}^4$, but with a different foliation of spacelike hypersurfaces orthogonal to a chosen time slicing. One may hope that it would be possible to find gravitational duals for {\sl all} boundary conditions that preserve 16 supercharges and, in particular, to identify individual duals for the specific boundary conditions that we discussed in detail in this paper. This would enable us to verify the claims we made in this paper in the large $K$ limit, and to shed more light on the remaining open issues.

\subsection{More on Enhanced Global Symmetries}

We argued (see also \cite{Beem:2013sza,Beem:2014zpa}) that four-dimensional ${\cal N}=2$ SCFTs cannot have enhanced global symmetries as one dials exactly marginal couplings, except in the situation that they are also combined with enhanced higher spin symmetries. It would be interesting to generalize this to other dimensions and other amounts of supersymmetry, and to study other similar constraints arising from the structure of the representation theory of the superconformal algebras \cite{Evtikhiev}. 

\subsection{Alternative Boundary Conditions in AdS}
 We argued that alternative boundary conditions must be possible for gauge fields on $AdS_5$, since in the field theory dual it is always possible to gauge global symmetries. But, we argued that since one has to ensure that the beta function of newly introduced gauge interactions vanish, such modified boundary conditions can never occur for weakly coupled gauge theories on $AdS_5$ space (which is where the boundary conditions have been classified), but only at strong coupling. It would be interesting to generalize this to other fields and other dimensions, to see whether alternative boundary conditions at strong coupling can arise there as well.

\section*{Acknowledgements}
We would like to thank Michael Douglas, Nadav Drukker, Jaume Gomis, Rajesh Gopakumar, Sergei Gukov, Marc Henneaux, Stefan Hohenegger, Amer Iqbal,  David Kutasov, Albion Lawrence, Shlomo Razamat, Joerg Teschner, Cumrun Vafa and Misha Vasiliev for useful discussions. We especially thank Zohar Komargodski and Yuji Tachikawa for many useful discussions and comments on a draft of this paper. The work of OA and MB was supported in part by an Israel Science Foundation center for excellence grant, by the Minerva foundation with funding from the Federal German Ministry for Education and Research, and by the I-CORE program of the Planning and Budgeting Committee and the Israel Science Foundation (grant number 1937/12). OA is the Samuel Sebba Professorial Chair of Pure and Applied Physics, and he is supported also by a Henri Gutwirth award from the Henri Gutwirth Fund for the Promotion of Research, and by the ISF within the ISF-UGC joint research program framework (grant no. 1200/14). SJR was supported in part by the National Research Foundation of Korea(NRF) grant funded by the Korea government(MSIP) through Seoul National University with grant numbers 2005-0093843, 2010-220-C00003 and 2012K2A1A9055280.

\appendix
\section{Conformal Structure of $AdS_d \times \mathbb{S}^1$}
\label{appendixA}

In this appendix, we explain how SCFTs can be defined consistently on the product space $AdS_d \times \mathbb{S}^1$ or its Euclidean counterpart $\mathbb{H}_d \times \mathbb{S}^1$ \footnote{A related situation arose in the context of the so-called supersymmetric Renyi entropy in three-dimensional conformal field theories and their holographic dual in terms of topological black holes \cite{Nishioka:2013haa,Huang:2014gca}.}. In global coordinates, the space is described by the metric
\be
\rmd s_{d+1}^2 = R_{AdS}^2 [\rmd \eta^2 + \sinh^2 (\eta) \rmd \Omega_{d-1}^2] + R_1^2 (\rmd \varphi)^2,
\ee
where
\be
0 \le \eta < \infty \quad \mbox{and} \quad 0 \le \varphi \le 2 \pi
\ee
and $R_{AdS}, R_1$ are the radii of curvature for $AdS_d$ and $\mathbb{S}^1$, respectively.

The above space is conformally equivalent to a branched sphere, $\mathbb{S}^{d+1}_q$ with deformation parameter $q \equiv (R_1/R_{AdS})$. Recall that a branched sphere is formed from a sphere by inserting conic singularities, where $(q-1)$ is the deformation parameter away from the round sphere
$\mathbb{S}^{d+1}$.
To see this, change the noncompact $\eta$ coordinate to a compact, angular coordinate $\theta$ by
\be
\sinh (\eta) = \cot (\theta) \quad \mbox{with} \quad 0 \le \theta \le {\pi \over 2}.
\ee
Note that the change of variable maps the $AdS_d$ asymptotic infinity to a point $\theta = 0$. Explicitly,
\be
\rmd s^2_{d+1} &=& {1 \over \sin^2 (\theta)} \left[ R_{AdS}^2 (\rmd \theta^2 + \cos^2 (\theta) \rmd \Omega_{d-1}^2)\right] + R_1^2 \rmd \varphi^2 \nonumber \\
&=& {1 \over \sin^2 (\theta)} R_{AdS}^2 \left[ \rmd \theta^2 + \cos^2 (\theta) \rmd \Omega_{d-1}^2 + q^2 \sin^2 (\theta) \rmd \varphi^2 \right].
\ee
We see that the expression inside the square bracket is precisely the line element of the branched sphere $\mathbb{S}^{d+1}_q$, where $q \rightarrow 1$ yields the line element of the round sphere $\mathbb{S}^{d+1}$ expressed with manifest $SO(d)$ invariance (this case is also conformally equivalent to flat space). In terms of embedding coordinates in $\mathbb{R}^{d+2}$, the branched sphere is defined by the hypersurface
\be
(X_1^2 + \cdots + X_d^2) + q^2 (X^2_{d+1} + X^2_{d+2}) = R^2.
\ee
The deformation has two branches:
\be
\mbox{Branch \ (1)}: \quad 0 \le q \le 1, \qquad \mbox{Branch \ (2)}: \quad 1 \le q < \infty.
\ee
For $d=2$ the two branches are related by a 'duality' transformation
\be
q \qquad \longleftrightarrow \qquad q^{-1}.
\ee

The round-sphere $\mathbb{S}^{d+1}$ admits Killing spinors. Upon the Weyl transformation and the coordinate transformation, they are mapped to Killing spinors on $AdS_d \times \mathbb{S}^1$ of equal radii. Once $q$ is turned away from 1, they cease to be Killing spinors of the branched sphere $\mathbb{S}^{d+1}_q$. To maintain them as Killing spinors, it is necessary to turn on non-geometric backgrounds that turn on the R-symmetry currents.

As $AdS_d \times \mathbb{S}^1$ and the branched sphere $\mathbb{S}^{d+1}_q$ are conformally equivalent, CFTs defined on these spaces are identical, with the exception of the conformal coupling of the scalar fields.

\section{${\cal N}=(2,0)$ theory on $AdS_5 \times \mathbb{S}^1$}
\label{twistedspectrum}

In this section, we review the Kaluza-Klein spectrum of the massless states (from the six-dimensional point of view) coming from the twisted sector of Type IIB string theory on $AdS_5 \times \mathbb{S}^5/\mathbb{Z}_K$, and its matching with the single-trace twisted sector operators of four-dimensional ${\cal N}=2$ superconformal quiver gauge theories \cite{Douglas:1996sw,Gukov:1998kk,Berenstein:2000hy}.

The massless states in the twisted sector are the same as those that arise on the `moduli space' of the six-dimensional ${\cal N}=(2, 0)$ $A_{K-1}$ theory \footnote{The analysis below is generalizable to any discrete subgroups $\Gamma \subset SU(2) \subset SO(4)$. In those cases, the number of six-dimensional tensor multiplets equals the number of non-trivial conjugacy classes of $\Gamma$. In particular, the generalization works straightforwardly for non-Abelian discrete subgroups of $SU(2)$.} on $AdS_5 \times \mathbb{S}^1$, namely $(K-1)$ tensor multiplets. The ${\cal N}=(2,0)$ tensor multiplet consists of a self-dual two-form potential, 4 complex spinors and 5 real scalars. In flat space they are singlet, spinor and vector representations of the $SO(5)$ R-symmetry, respectively. On $AdS_5 \times \mathbb{S}^5/\mathbb{Z}_K$, we have instead the $SU(2)_R \times U(1)_R$ R-symmetry of the ${\cal N}=2$ $AdS_5$ superalgebra. On $AdS_5\times \mathbb{S}^1$, the massless states arising from the self-dual two-form potential are an $SU(2)_R$-singlet vector potential on $AdS_5$. The 4 complex spinors are decomposed to two $SU(2)_R$-doublet spinors, with $U(1)_R$ charges $\pm 1/2$. The 5 real scalar fields are decomposed to a triplet and two singlets of $SU(2)_R$, with $U(1)_R$ charges $0, \pm 1$.

The six-dimensional ${\cal N}=(2,0)$ Poincar\' e supersymmetry is described by the $OSp(8^* \vert 4)$ superalgebra, whose minimal unitary supermultiplet corresponds to the tensor multiplet. The five-dimensional ${\cal N}=2$ AdS supersymmetry is described by the $SU(2, 2 \vert 2)$ superalgebra, whose minimal unitary supermultiplet corresponds to the vector multiplet. The super-generators of the $SU(2,2|2)$ superalgebra take the form
\be
SU(2,2 | 2): \qquad
\left[ \begin{array}{c c c|c c c}
& & & & & \\
\ \ & L^{mn} & \ \ & \ \ & Q^{\alpha a} & \ \ \\
& & & & & \\
\hline
& & & & & \\
\ \ & \overline{Q}^\beta_b & \ \ & \ \ & J_{ab} & \ \ \\
& & & & &
\end{array}
\right]
\ee
These generators can be embedded into the $OSp(8^*|4)$ superalgebra by compactifying over $SO(2) \subset SO^*(8)$:
\be
OSp(8^*\vert 4): \qquad
\left[ \begin{array}{ccc|ccc}
&&&&& \\
\ \ & L^{MN} & \ \ & \ \  & Q^{\mu A}& \ \  \\
\hline
\ \  
&&&&& \\
&\overline{Q}^\nu_B & \ \ & \ \ & J_{AB} & \ \ \\
&&&&&
\end{array}
\right] =
\left[ \begin{array}{ccc|ccc}
\ \  & L^{mn} & - & Q^{\alpha a}& - &\ \  \\
\ \  & - & R_{SO(2)} & - & - &\ \ \\
\hline
&&&&& \\
\ \  &\overline{Q}^\beta_b & - & J_{ab} & - &\ \ \\
&-&-&-&-&
\end{array}
\right]
\ee
Recall that the extra $U(1)_R$ symmetry of the $SU(2,2|2)$ superalgebra descends from the isometry of the $\mathbb{S}^1$ compactification, so it is embeddable to the $SO^*(8) \simeq Spin(6,2)$ bosonic subalgebra of the $OSp(8^*|4)$ superalgebra.

We now make a Kaluza-Klein compactification on $\mathbb{S}^1$ for the bosonic massless fields of the twisted sector. As recalled in section \ref{basicsec}, the $U(1)_R$ part of the R-symmetry of the ${\cal N}=2$ $AdS_5$ super-algebra includes the isometry of this $\mathbb{S}^1$. As such, Kaluza-Klein excitations in the same multiplet are multiply charged under the $U(1)_R$. The mass spectra of the Kaluza-Klein states depend on two sources. The first source is the harmonics on $\mathbb{S}^1$. For bosonic fields, its contribution is integrally quantized and yields a universal spectrum $m^2 = (\mathbb{Z})^2$, measured in units of $1/R_{S^1}^2$ of the $\mathbb{S}^1$. The second source is background-induced couplings when the six-dimensional ${\cal N}=(2,0)$ theory is put on $AdS_5$ with non-trivial metric curvature and Ramond-Ramond 5-form flux. The bosonic fields are:
\begin{itemize}
\item $SU(2)_R$ triplet scalar fields ${\boldsymbol \Phi}_a = (\phi_1, \phi_2, \phi_3)_a, \ (a=1, \cdots, K-1)$: For $A_{K-1}$, there are $(K-1)$ sets of $SU(2)_R$ triplet scalar fields. Geometrically, they represent the 3 blow-up modes of each of the $(K-1)$ 2-cycles that one can blow up from $\mathbb{C}^2/\mathbb{Z}_K$. 
Globally, these cycles are embedded on a hypersurface inside $\mathbb{S}^5/\mathbb{Z}_K$. Since the latter is positively curved with curvature scalar $4$ (in units of $1/R_{S^1}^2$), the triplet scalar fields acquire a tachyonic mass-squared $-4$ on $AdS_5 \times \mathbb{S}^1$. Upon Kaluza-Klein compactification on $\mathbb{S}^1$, these scalar fields thus acquire the mass-squared spectrum on $AdS_5$ :
\be
\mbox{$SU(2)_R$ \ triplet scalar fields}: \qquad m^2 = n^2 -4.
\label{triplet}
\ee
Here, $n$ labels the $U(1)_R$ charges, so $ n >0$ states and $n <0$ states are complex conjugates of each other, giving
complex scalars with R-charge $n$. 
The $n=0$ state is real and corresponds to the zero-mode on $\mathbb{S}^1$ with $m^2 = -4$. The dimensions of the
correponding operators ${\cal O}^{(n)}$ are $\Delta = 2 + |n|$.

\item $SU(2)_R$ singlet scalar fields $(\phi_4, \phi_5)_a, \ (a=1, \cdots, K-1)$: The two scalar fields are periods of the two two-form potentials in Type IIB string theory over the blow-up homology 2-cycles $\Sigma_a \ (a = 1, \cdots, K-1)$. These 2-form potentials interact with the background 5-form field strength $G_5$ that permeates $AdS_5$ through the parity-odd Chern-Simons term $ 4 G_5 \wedge [B_2 \wedge \rmd C_2 - C_2 \wedge \rmd B_2]$ in ten-dimensional Type IIB supergravity. The non-trivial second cohomology group of the orbifold is generated by $(K-1)$ anti-self-dual normalizable harmonic 2-forms. Dimensionally reducing over the orbifold and Kaluza-Klein compactifying over the $\mathbb{S}^1$, this Chern-Simons coupling gives rise to off-diagonal mass terms $4 n (\phi_{4,n} \phi_{5,-n} + \phi_{4,-n} \phi_{5,n})$ for the $n$-th harmonic excitations. Diagonalizing, these scalar fields exhibit the mass-squared spectrum on $AdS_5$:
\be
\mbox{$SU(2)_R$ \ singlet scalar fields}: \qquad 
&& m^2 = n^2 + 4 n, \nonumber \\                                              
&& m^2 = n^2 - 4n.
\label{complexsinglet}
\ee
Again, $n$ labels the $U(1)_R$ charge. 
The mode from the first tower with R-charge $n$ is the complex conjugate of the mode from the second tower with R-charge $(-n)$, which
has the same mass. Thus for every $n$ we get a complex scalar with R-charge $n$, corresponding to an operator of dimension
$\Delta = 2 + |n+2|$. Equivalently, for every $n > 0$ we have two
complex scalars of R-charge $n$, whose corresponding operators ${\cal O}_+^{(n)}$ and ${\cal O}_-^{(n)}$ have scaling dimensions $\Delta = 2 + |n \pm 2|$, respectively. For $n=0$ we have a single
massless complex scalar, corresponding to an operator ${\cal O}_+^{(0)}$ with $\Delta=4$.


\item $SU(2)_R$ singlet self-dual 2-form fields $B^{(2)}_{a}, \ (a = 1, \cdots, K-1)$: These fields are not affected by the curvature and background flux of $AdS_5$. So, the Kaluza-Klein compactification yields tensor fields whose mass-squared spectrum on $AdS_5$ is given by
\be
\mbox{$SU(2)_R$ singlet tensor fields}: \qquad m^2 = n^2.
\label{neutralsinglet}
\ee
Here again, $n$ labels the $U(1)_R$ charge. For $n \neq 0$ these are massive 2-form fields, and the $n>0$ modes are complex
conjugates of the $n<0$ modes, while for $n=0$ we should dualize
the massless 2-form to a real massless vector field. For $n \neq 0$ the corresponding two-form operators ${\cal O}_{\mu \nu}^{(n)}$ 
have dimension $2+|n|$, while for $n=0$ the operator ${\cal O}_{\mu}^{(0)}$ is a conserved current of dimension $\Delta=3$.

\end{itemize}\emph{}

We can match these spectra with the BPS operators in the `twisted sector' of the four-dimensional ${\cal N}=2$ SCFT with $\widehat{A}_{K-1}$ quiver gauge group. At the orbifold point, they are organized according to the representation of the $\mathbb{Z}_K$ quantum symmetry. We shall denote the $K$-th root of the unity as $\omega$, $\omega^K = 1$. Away from the orbifold point there are some deformations of these operators which we will not write down explicitly. 

The four-dimensional ${\cal N}=2$ quiver SCFT has $K$ vector multiplets $(\varphi, \lambda, F_{\mu \nu})_a$ in adjoint representations,
with $U(1)_R$-charges $(1,1/2,0)$, and with $\lambda$ a doublet of $SU(2)_R$, and $K$ hypermultiplets $(\chi, H,  \widetilde{\chi})_{a, a+1}$ in bi-fundamental representations,
with $U(1)_R$ charges $(1/2,0,-1/2)$, and with $H$ a complex doublet of $SU(2)_R$. Representatives of the BPS operators corresponding to most of the Kaluza-Klein towers described above are then
\be
{\boldsymbol{ \cal O}}_a^{(n)}  & =& \sum_{b = 0}^{K-1} 
(\omega^a)^b \ 
\mbox{Tr}_b \big({\lambda} {\boldsymbol{\vec{\sigma}}} 
{\lambda} \ {\varphi}^{n-1} \big)_b 
\hskip2.2cm n \ge 1, \nonumber \\
{\cal O}_{+,a}^{(n)} & =& \sum_{b=0}^{K-1} (\omega^a)^b 
\ \mbox{Tr}_b \left( (F^2 + i F \wedge F) {\varphi}^{n} \right)_b \hskip0.9cm n \ge 0, \nonumber \\
{\cal O}_{-,a}^{(n)} & =&  \sum_{b=0}^{K-1} (\omega^a)^b
\ \mbox{Tr}_b \left({\varphi}^{n} \right)_b 
\hskip3.5cm n \ge 2, \nonumber \\
{\cal O}_{\mu \nu,a}^{(n)} & = & \sum_{b=0}^{K-1} (\omega^a)^b \
\mbox{Tr}_b \left(F_{\mu \nu} {\varphi}^n \right)_b 
\hskip3.0cm n \ge 1,
\label{ftops}
\ee
where $b$ labels the $SU(N)$ group inside $SU(N)^K$, and $a = 1, \cdots, K-1$ labels the $(K-1)$ different twisted sectors. Their $SU(2)_R$ and $U(1)_R$ quantum numbers match precisely with the corresponding Kaluza-Klein 
momentum states of (\ref{triplet}), (\ref{complexsinglet}) and (\ref{neutralsinglet}), respectively. Their $U(1)_R$ charges are all $n$, while their scaling dimensions are $(n+2), (n+4), (n), (n+2)$, respectively. The modes above with $U(1)_R$ charges $(n,n-1,n+1,n)$ are the bosonic operators sitting in a specific multiplet of the five-dimensional supersymmetry in the bulk. 

We are still missing a few low-dimension operators which have special forms. In the first line we are missing the real
scalars with $n=0$, which may be written as
%
\be
{\cal O}_a^{(0)} &=& \sum_{b=0}^{K-1} (\omega^a)^b \ \mbox{Tr}_b (H {\boldsymbol {\vec \sigma}} \overline{H})_b,
\ee
where the trace here includes contributions from both hypermultiplets coupled to the $b$'th
$SU(N)$ group.
In the third line we are missing a complex operator with $n=1$, which may be written as
\be
{\cal O}_{-,a}^{(1)}&=& \sum_{b=0}^{K-1} (\omega^a)^b  
\ \mbox{Tr}_b (\lambda^2)_b.
\ee
The missing operators in the fourth line with $n=0$ are precisely the `relative' $U(1)^{K-1}$ currents acting on the
hypermultiplets. Note that if our gauge group was $U(N)^K$ we would have the operators in the fourth line of
\eqref{ftops}, but since our gauge group is $SU(N)^K$, we have instead the $U(1)$ currents that the $U(1)^K$
gauge fields would have coupled to; in the $U(N)^K$ theory they are given by acting by $\partial^{\mu}$ on
the last line of \eqref{ftops} and using the equation of motion. This is equivalent to the statement that in the holographic dual, the Kaluza-Klein reduction of the self-dual tensor yields a vector multiplet for the massless modes, as opposed to tensor multiplets for the massive modes. 
%
The dimensions of all these operators match with the spectrum in the bulk described above.


\end{document}